\documentclass[journal]{IEEEtran}
\usepackage{textgreek}
\usepackage{multirow}
\usepackage{booktabs}
\usepackage{subcaption}
\usepackage{dblfloatfix}
\usepackage[table,xcdraw]{xcolor}
\usepackage{threeparttable}
\usepackage{fancyhdr}

\usepackage{cite}

\ifCLASSINFOpdf
  \usepackage[pdftex]{graphicx}
  \graphicspath{{./Figs/}}
\else
  \usepackage[dvips]{graphicx}
\fi
\usepackage{amsmath}
\usepackage{scalerel}
\usepackage{tikz}
\usetikzlibrary{svg.path}
\definecolor{orcidlogocol}{HTML}{A6CE39}
\tikzset{
    orcidlogo/.pic={
        \fill[orcidlogocol] svg{M256,128c0,70.7-57.3,128-128,128C57.3,256,0,198.7,0,128C0,57.3,57.3,0,128,0C198.7,0,256,57.3,256,128z};
        \fill[white] svg{M86.3,186.2H70.9V79.1h15.4v48.4V186.2z}
        svg{M108.9,79.1h41.6c39.6,0,57,28.3,57,53.6c0,27.5-21.5,53.6-56.8,53.6h-41.8V79.1z M124.3,172.4h24.5c34.9,0,42.9-26.5,42.9-39.7c0-21.5-13.7-39.7-43.7-39.7h-23.7V172.4z}
        svg{M88.7,56.8c0,5.5-4.5,10.1-10.1,10.1c-5.6,0-10.1-4.6-10.1-10.1c0-5.6,4.5-10.1,10.1-10.1C84.2,46.7,88.7,51.3,88.7,56.8z};
    }
}
\newcommand\orcidicon[1]{
\href{https://orcid.org/#1}
{\mbox{\scalerel*{
                \begin{tikzpicture}[yscale=-1,transform shape]
                \pic{orcidlogo};
                \end{tikzpicture}
            }{|}}}}
            
\usepackage{hyperref} 
\hypersetup{hidelinks}

\usepackage{soul}
\newcommand{\deltext}[1]{\relax}
\newcommand{\addtext}[1]{#1}
\newcommand{\delRev}[1]{\relax}
\newcommand{\addRev}[1]{#1}

\begin{document}
%
\title{An Energy-efficient Capacitive-\deltext{Memristive}\addtext{RRAM} Content Addressable Memory}
%
%
%

\author{Yihan Pan$^{\textsuperscript{\orcidicon{0000-0002-2666-5540}}}$\,~\IEEEmembership{Graduate Student Member,~IEEE, }%
Adrian Wheeldon$^{\textsuperscript{\orcidicon{0000-0003-4672-5990}}}$\,~\IEEEmembership{Member,~IEEE, }%
\newline Mohammed Mughal$^{\textsuperscript{\orcidicon{0000-0002-1061-6576}}}$\,~\IEEEmembership{Member,~IEEE, }%
Shady Agwa,~\IEEEmembership{Member,~IEEE, }%
\newline Themis Prodromakis$^{\textsuperscript{\orcidicon{0000-0002-6267-6909}}}$\,~\IEEEmembership{Senior Member,~IEEE, }%
 and Alexantrou Serb$^{\textsuperscript{\orcidicon{0000-0002-8034-2398}}}$\,~\IEEEmembership{Senior Member,~IEEE, }%
\thanks{Y. Pan, A. Wheeldon, M. Mughal, S. Agwa, T. Prodromakis and A. Serb are with the Centre for Electronics Frontiers, Institute for Integrated Micro and Nano Systems, School of Engineering, University of Edinburgh, EH9 3LA, UK. (emails: yihan.pan@ed.ac.uk; adrian.wheeldon@ed.ac.uk; mmughal@exseed.ed.ac.uk; shady.agwa@ed.ac.uk; t.prodromakis@ed.ac.uk; aserb@ed.ac.uk).

}%
}

%
%

\renewcommand{\headrulewidth}{0pt}
\pagestyle{fancy}
\fancyhf{}
\fancyhf[HC]{This article has been accepted for publication in IEEE Transactions on Circuits and Systems--I: Regular Papers. This is the author's version which has not been fully edited and content may change prior to final publication. Citation information: DOI 10.1109/TCASI.2024.3451707}
\fancyfoot{} 
\fancyfoot[FC]{Personal use is permitted, but republication/redistribution requires IEEE permission. See https://www.ieee.org/publications/rights/index.html for more information.}

\markboth{IEEE TRANSACTIONS ON CIRCUITS AND SYSTEMS—I: REGULAR PAPERS,~Vol.~X, No.~X, Month~Year}%
{Yihan \MakeLowercase{\textit{et al.}}: An Energy-efficient Capacitive-Memristive Content Addressable Memory}
%

\IEEEpubid{0000--0000/00\$00.00~\copyright~2024 IEEE}


\maketitle

\begin{abstract}
Content addressable memory is popular in \deltext{the field of} intelligent computing systems \deltext{with its searching nature}\addtext{as it allows parallel content-searching in memory}. Emerging CAMs show a promising increase in \deltext{pixel}\addtext{bitcell} density and a decrease in power consumption than pure CMOS solutions. This article introduced an energy-efficient 3T1R1C TCAM cooperating with capacitor dividers and RRAM devices. The RRAM as a storage element also acts as a switch to the capacitor divider while searching for content. CAM cells benefit from working parallel in an array structure. We implemented a 64x64 array and digital controllers to perform with an internal built-in clock frequency of 875MHz. Both data searches and reads take \deltext{3x}\addtext{three} clock cycles. Its worst average energy for data match is reported to be 1.71fJ/bit-search and the worst average energy for data miss is found at 4.69fJ/bit-search. The prototype is simulated and fabricated in 0.18um technology with in-lab RRAM post-processing. Such memory explores the charge domain searching mechanism and can be applied to data centers that are power-hungry.
\end{abstract}

\begin{IEEEkeywords}
Content addressable memory(CAM), capacitive-divider, resistive random-access memory(RRAM)
\end{IEEEkeywords}

%
\IEEEpeerreviewmaketitle

\section{Introduction}
\IEEEPARstart{T}{raditional} memory structures such as SRAM and DRAM are address-addressable only. With the development of intelligent systems, address-addressable-only memories limit computing efficiency in terms of data processing and transferring. Advanced intelligent systems require not only memory storage but also the ability to locate memory addresses by certain content\cite{karam2015emerging}. We call this type of memory content addressable memory (CAM), also known as associative memory. The search-by-content manner of CAMs enables data visible to the system and accelerates information processes to focus on their location of interests without other extra data movement compared to the von Neumann architecture. Therefore, the ability of bit storage and comparison has made CAM attractive to the field of network routers \cite{ooka2014networkrouter}, packet classification \cite{liu2016packet}, neuromorphic computing \cite{ni2019ferroelectric}, reconfigurable computing \cite{paul2008reconfigurable}, pattern recognition \cite{graves2020memory4t2r}, and data compression \cite{lin2000datacomp}. 

Conventional CAM in CMOS technology normally uses SRAM as memory elements to store data and its data comparison is implemented by NAND-based \cite{NANDNOR2011} or NOR-based \cite{NOR2013} topologies. A basic binary CAM typically has 10T (transistors) whereas it also can be implemented into a ternary CAM that takes two SRAM together to acquire an additional state called `don't care' in 16T \cite{pagiamtzis2006content}. CAM cell allows parallel searching and they normally been built with memory arrays. Studies have looked for a denser and more robust solution with less power consumption in the SRAM-based design \cite{lin20037tcmos}. Although the CAM memory is desired because of its high operational speed and searching function, its density increases its cost of implementation more than the other standard memories in CMOS. Thus, emerging memories beyond CMOS technologies have been taken into the field to maintain a high throughput rate with a reduction in area \cite{karam2015emerging}.

Emerging \addtext{memristive} devices are treated as canonically non-volatile and their characteristics depend on the properties of their fabrication materials or structures. They are categorized by their operation mechanisms such as Resistive Random Access Memory (RRAM) \cite{prodromakis2011RRAM}, Phase Change Memory (PCM) \cite{burr2010phase}, Ferroelectric RAM (FeRAM) \cite{lue2003FeRAM}, and magnetic RAM (MRAM) \cite{dong2008MRAM}. Those devices have a common factor that they can be altered to different states, leading to a suitable technique for memory. As a result, the bit storage for CAM with emerging technologies is achieved by the device itself instead of two SRAMs (12T). At the same time, the bit comparison utilizes the differences between states to distinguish the stored bit on the device. In this work, we are focused on RRAM whose resistivity can be altered to different levels. The Metal-Insulator-Metal (MIM) structure of RRAM changes its resistance by redox processes, coupling to ion-migration effects \cite{Waser2009}. In other words, the device resistance can be switched to a low resistance state (LRS) or a high resistance state (HRS) by applying set and reset threshold pulses respectively. Some material structure also allows middle resistance states to form the RRAM a multi-bit memory cell \cite{stathopoulos2017multibit}. 

\begin{figure*}[htb]
    \begin{minipage}[t]{.48\textwidth}
        \centering
        \includegraphics[width=\textwidth]{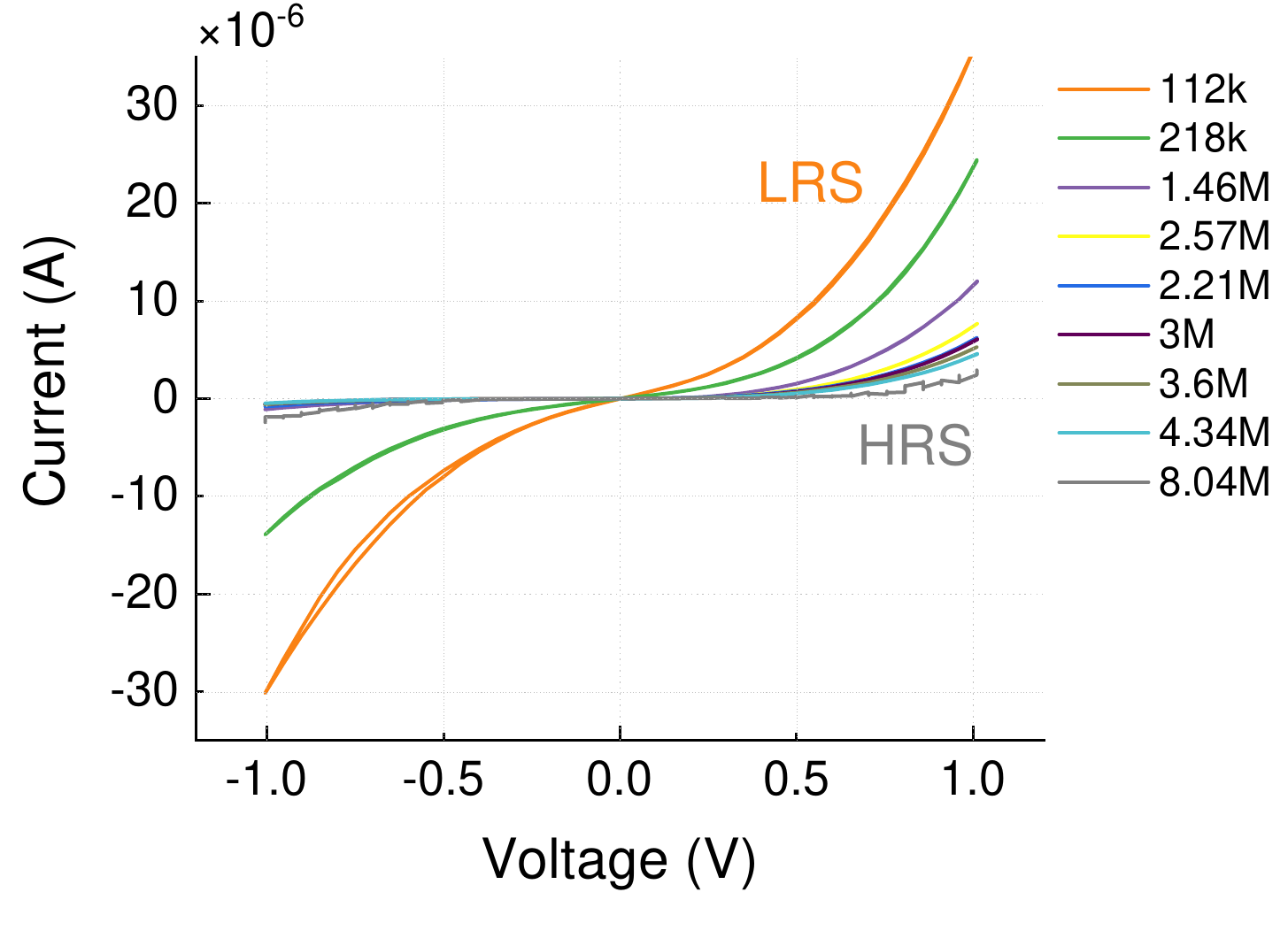}
        \subcaption{}\label{fig:static_IV_full}
    \end{minipage}
    \hfill
    \begin{minipage}[t]{.48\textwidth}     
        \centering
        \includegraphics[width=\textwidth]{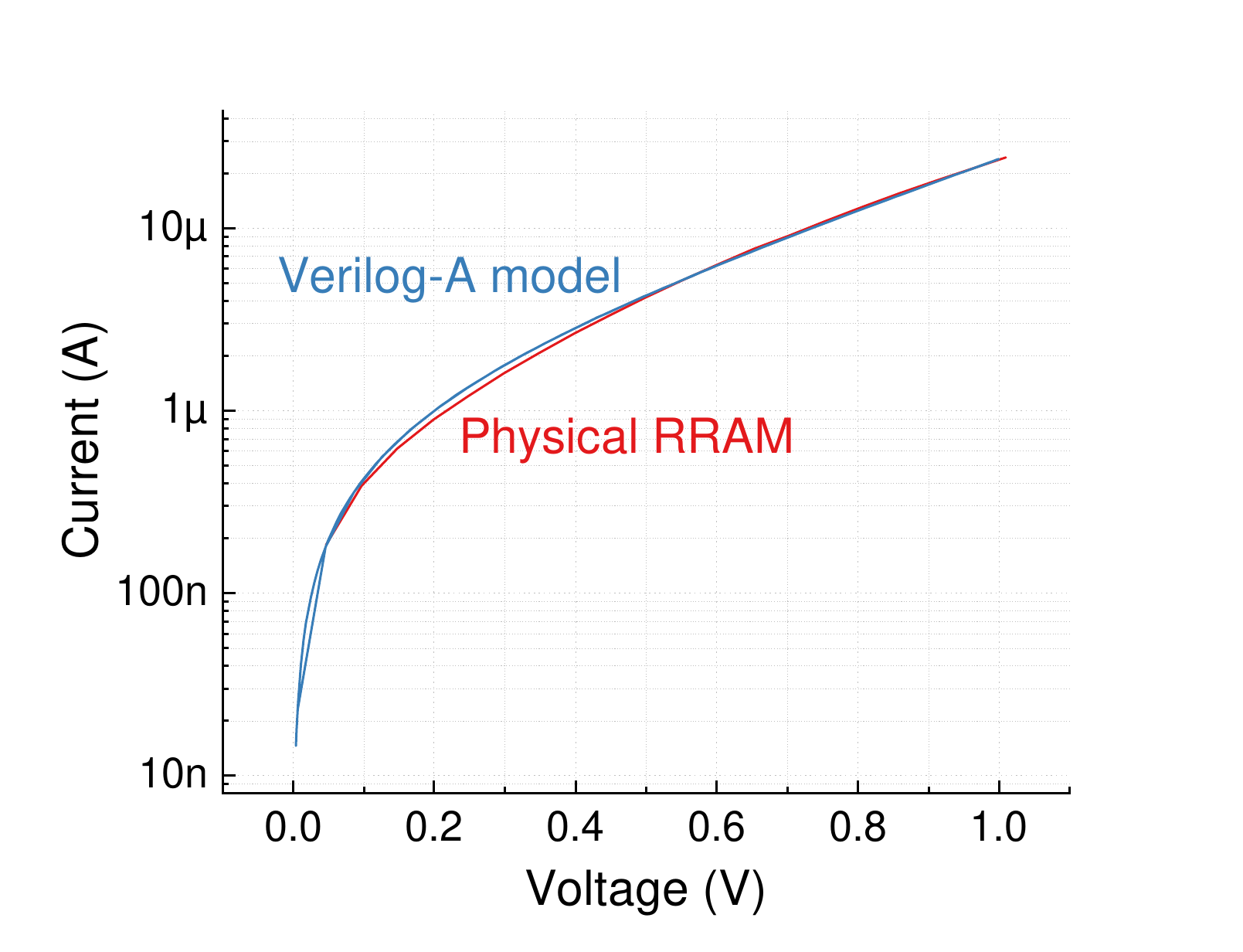}
        \subcaption{}\label{fig:static_IV_fit}
    \end{minipage}
    \caption{(a) Measured RRAM static IV characteristics at different resistance levels. Resistance values in the legend are quoted in Ohms and for a read-out voltage of 0.2V. (b) Pristine RRAM static IV comparison with its fit line at 218k\textOmega \space in the logarithmic domain (positive direction).}
\end{figure*}

\IEEEpubidadjcol
Existing non-volatile CAMs are proposed with a much dense area. 
\addRev{A significant portion of their energy consumption is attributed to DC power dissipation in the cells and match-line energy. However, they have a lower normalized search energy compared to 16T CMOS TCAM\cite{2.5T}.}
The RC-XNOR-Z in \cite{pan2021} uses RRAM for storage and XNOR gate for evaluation. The non-volatile device can be also used for evaluation such as the 4T2R RRAM-based TCAM proposed in \cite{chang2016reram4t2r}\addRev{, 3T1R TCAM \cite{chang20173t1r}, 4T2R cell from Hewlett Packard lab\cite{graves2019_4t2r}, and 2T2R structure based on PCM \cite{li20132t2rIBM}. They imply}\delRev{. It implies an} RC delay \addRev{for evaluation} but suffer\delRev{s} from a long recovery time for high RRAM resistive states. 
\delRev{The same group also researched a 3T1R TCAM [21] further to improve the performance. However, the 3T1R solution works as a resistor divider which has a}
\addRev{The} direct current path \addRev{in those cells consume}\delRev{, consuming} a significant portion of energy inefficiently, waiting for RC stabilization. 
\delRev{Hewlett Packard lab developed a 4T2R cell in 2019 [22] with improved energy efficiency but longer search latency. IBM contributes to a 2T2R structure based on PCM[23] with a 2-bit encoding scheme but with two transistors stressing on the same match-line per bitcell.} 
The memory development in the charge domain is also designed with FeFET in 2021 \cite{yin2021CapFeFET}. Analogue CAM has also been developed by \cite{li2020analog} where the multilevel of RRAM resistive states shows the possibility for complex searches. Overall, the RRAM-based TCAM shows promising performance in search delay and power among different types of TCAMs \cite{karam2015emerging}. It is essential to find improved power efficiency solutions with a high throughput rate at low cost and high reliability for CAM designs\addRev{, especially under all-mismatch cases where the cell DC energy occupies nearly half of the total energy consumption of non-volatile TCAMs \cite{2.5T}.}

In this paper, we propose a capacitive-based RRAM CAM with energy-efficient solutions operated with an 875 MHz clock. \deltext{The pixel operates in the charge domain with no direct current path. It offers both address-addressable read and content-addressable search with lower energy consumption compared to the state-of-the-art.}\addtext{Our proposed work applies capacitors to operate searches in the charge domain to eliminate the in-efficient energy consumption of direct current path in most of the resistive-network CAM cells\delRev{, especially in the 3T1R design,} without adding extra transistor and RRAM to the bitcell\delRev{like the 4T2R design}. \addRev{The introduction of capacitor in cell eliminates the DC portion of power consumption compared to the resistive designs.} Apart from searching content within the memory (content-addressable read), it also offers memory access in the traditional way like SRAM/DRAM to read the memory by a given address (address-addressable read).} In section II, we first introduce an RRAM model built \deltext{based}\addtext{from measured data} on our own fabricated devices. Then, the \deltext{pixel}\addtext{CAM cell} and system performance are highlighted in section III. Section IV provides the analysis of the experimental measurements and follows with further discussions and a conclusion at the end. 

\section{RRAM model}
In this work, the in-house fabricated non-volatile RRAM with a structure of Au/TiO2/Pt is applied to the CAM \deltext{pixel}\addtext{cell}. We model our RRAM devices based on the modeling approach in \cite{Messaris2018_verilogA}, where we have modified the equations for the IV characteristics as shown below. As in the original modeling work, device model parameters are directly derived from measured static IV characteristics of Au/TiO2/Pt devices \cite{Tom2022RRAM} (i.e. it is a data-driven model), as shown in \deltext{f}\addtext{F}ig. \ref{fig:static_IV_full}. Here, we assumed that the resistive states of the device do not change during any read and search operations unless the devices are actively programmed. We fit experimental IV data into the following exponential equation:
\begin{equation} \label{equ:model}
    i(v)=\begin{cases}a_p(1/RS)(1-exp(-b_pv)) & v > 0\\a_n(1/RS)(1-exp(-b_nv)) & v < 0\end{cases}
\end{equation}
where $a_{p,n}$ and $b_{p,n}$ are fitting parameters at the resistance RS. A comparison example between a real device and its fitted model in \deltext{f}\addtext{F}ig. \ref{fig:static_IV_fit} proves its accuracy. The exponential relation indicates a similarity between RRAM devices and diodes.

\begin{figure*}[!t]
    \begin{minipage}[t]{.35\textwidth}
        \centering
        \includegraphics[width=\textwidth]{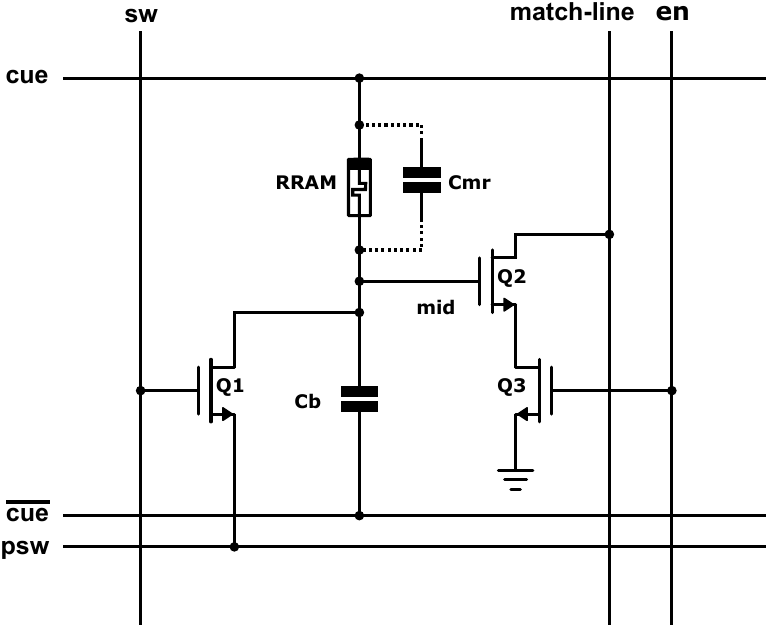}
        \subcaption{}\label{fig:Pixel}
    \end{minipage}
    \hfill
    \begin{minipage}[t]{.3\textwidth}     
        \centering
        \includegraphics[width=0.88\textwidth]{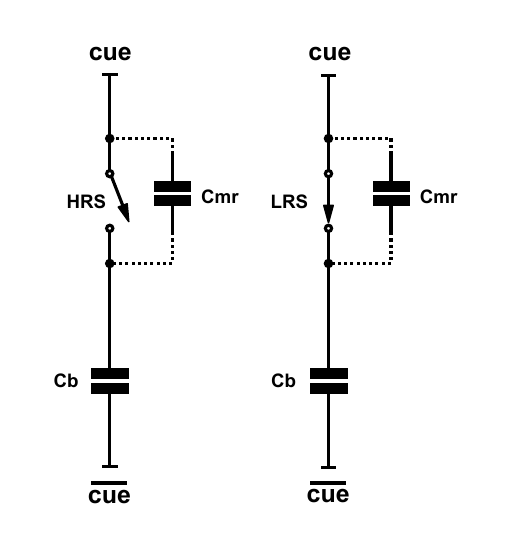}
        \subcaption{}\label{fig:Mid_Branch}
    \end{minipage}
    \hfill
    \begin{minipage}[t]{.33\textwidth}
        \centering
        \includegraphics[width=\textwidth]{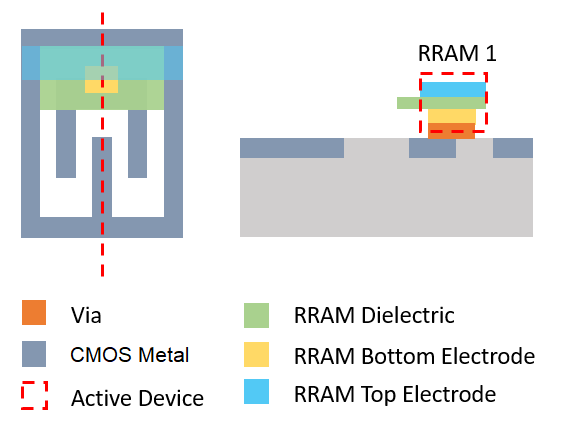}
        \subcaption{}\label{fig:RRAM_Position}
    \end{minipage}  
    \caption{(a) \deltext{Pixel}\addtext{3T1R1C CAM cell} schematic. C{\footnotesize mr} is the RRAM-introduced parasitic capacitance that forms a capacitor divider with the physical capacitance C{\footnotesize b}. Q2-3 build a double-gated transistor. (b) RRAM-C branch equivalent concept shows the memristor acting as a switch. (c) The arrangement of post-processed RRAM \deltext{in pixels}\addtext{per CAM cell}. RRAM is placed at the top of the transistors and the capacitors.}
\end{figure*}

\section{System Overview}
The system constructs a 64x64 CAM array with RRAM \addtext{in-memory storing and computing data. }\deltext{in-pixel, it is}\addtext{Each CAM cell is} capable of performing content-addressable read, address-addressable read, and write operations. \addtext{The system implementation and layout are also provided in the following subsections.}

\subsection{\addtext{3T1R1C CAM cell}}
Fig. \ref{fig:Pixel} shows the TCAM \deltext{pixel}\addtext{cell} schematic in this work.\deltext{To maximize cell density, only nMOS transistors are used.} It has a 3T1R1C structure which consists of three transistors, an RRAM, and a capacitor C{\footnotesize b}. \addtext{To maximize cell density, only nMOS transistors are used.}\deltext{1-bit of data is stored in the RRAM represented by different resistive states.} \addtext{The RRAM stores 1-bit binary data in different resistive states. In this work, RRAM in HRS represents binary `1', and LRS means that binary `0' is stored.} \addtext{The introduction of RRAM also implies parasitic capacitance (shown as the dotted capacitance C{\footnotesize mr} in Fig. \ref{fig:Pixel}). It is}\deltext{The dotted capacitance C{\footnotesize mr} is the parasitic capacitance} introduced by the RRAM MIM structure, whose size is determined by the area of the fabricated RRAM. \addtext{This parasitic capacitance C{\footnotesize mr} forms a capacitive divider with another physical capacitor (called the bottom capacitor C{\footnotesize b} in Fig. \ref{fig:Pixel}). In this work, t}\deltext{T}he bottom capacitor C{\footnotesize b} is \deltext{chosen as}\addtext{implemented with }a Metal-Oxide-Metal (MOM) \deltext{capacitor}\addtext{structure} which uses inter-digitated fingers, customized to a smaller minimum capacitor than the foundry's standard cells (different choices of capacitor implementation may be warranted in different technologies). 

We highlight the physical\deltext{, in-pixel} RRAM position in \addtext{F}\deltext{f}ig. \ref{fig:RRAM_Position} where the active device is placed above the CMOS \deltext{metal}layer\addtext{s}. \addtext{For clarity, t}\deltext{T}he graph only shows the topmost CMOS metallization layer \deltext{of C{\footnotesize b} for clarity}\addtext{that is used to customize the bottom physical capacitor C{\footnotesize b}}. The \deltext{pixel} transistors and other \addtext{CMOS} layers are hidden below this metal layer \addtext{in grey}. The cross-section view \addtext{(on the right of Fig. \ref{fig:RRAM_Position})} is obtained from the \addtext{red} cut-line dissecting the top view \addtext{(on the left)}. This RRAM device has an area of 0.35\textmu m x 0.35\textmu m and its capacitance is approximated to 2.2fF based on measurements of physical RRAM devices.

The \addtext{CAM cell}\deltext{pixel} supports three operations: content-addressable read (CAR), address-addressable read (AAR), and write (WRT). The transistor Q1 is reused for both WRT and AAR operations, whereas transistors Q2-3 are only activated for CAR. \addtext{In this work, the }RRAM LRS is defined at 112k\textOmega \space and HRS at 8.04M\textOmega \space as they are reasonably easy to achieve in our practical RRAM\deltext{ and for which we have specific, measured models that we will be using throughout the rest of this work}. The circuit is operated with a primary 1.8V supply voltage V{\footnotesize DD} and an adjustable secondary supply voltage V{\footnotesize SEC}, here designed to range within [1, 1.4]V. V{\footnotesize SEC} is applied to cue/$\overline{\text{cue}}$ and pre-charge signals only.

\subsection*{Content-Addressable Read}
\begin{table}[!t]
\caption{3T1R1C voltage levels at different stages}
\label{tab:3t1r1c_voltage}
\centering
\begin{tabular}{@{}ccccc@{}}
\toprule
\textbf{Cue} & \textbf{Stored data} & \textbf{RRAM} & \textbf{V{\footnotesize mid}} & \textbf{Match-line} \\ \midrule
\multirow{2}{*}{1} & 1 & HRS & Low  & High \\ \cmidrule(l){2-5} 
                   & 0 & LRS & High & Low  \\ \midrule
\multirow{2}{*}{0} & 1 & HRS & High & Low  \\ \cmidrule(l){2-5}  
                   & 0 & LRS & Low  & High \\ \midrule 
\multirow{2}{*}{X} & 1 & HRS & Low  & High \\ \cmidrule(l){2-5} 
                   & 0 & LRS & Low & High  \\ \bottomrule
\end{tabular}
\end{table}

Content-addressable read can be operated in parallel throughout the entire array. The \addtext{searching} data \deltext{`1' is searching for (the ``cue"), }is input to cue and its complementary data to $\overline{\text{cue}}$ in \addtext{F}\deltext{f}ig. \ref{fig:Pixel}: when searching for a \addtext{binary} `1', we set cue to V{\footnotesize SEC} and $\overline{\text{cue}}$ to G{\footnotesize ND} and vice versa for a `0'. Additionally, if both cue and $\overline{\text{cue}}$ are set to G{\footnotesize ND}, the system performs a ``don't care" search (`X'). \addtext{Transistor} Q1 is turned off during the entire \addtext{CAR} search operation. The \addtext{resistive} state of the RRAM modulates the effective capacitive divider ratio by either by-passing C{\footnotesize mr} or not. \deltext{The divider mid-point node 'mid' directly drives the gate of Q2.} This is illustrated by the equivalent switched capacitor \deltext{network}\addtext{concept} shown in \addtext{F}\deltext{f}ig. \ref{fig:Mid_Branch}. When the RRAM is at HRS in M\textOmega, the RRAM behaves as an open switch, and the C{\footnotesize b}-C{\footnotesize mr} divider functions normally. When the RRAM is at LRS, \addtext{the voltage level at the divider mid-point }V{\footnotesize mid} directly follows cue (in the limit). In practice, because the LRS used is fairly high, we obtain effective capacitive divider modulation by shifting the RRAM-C{\footnotesize mr} RC constant. This is especially important at high speeds (\deltext{100s}\addtext{in hundreds} of MHz). At the same time, the behavior of V{\footnotesize mid} also depends on the cue/$\overline{\text{cue}}$ values \deltext{ resulting in t}\addtext{.T}able \ref{tab:3t1r1c_voltage} \addtext{lists the possible combinations of the searching cue and the stored data with the resulted V{\footnotesize mid} and match-line outcomes.} When the cue does not match the stored data (`\addtext{data-}miss'), V{\footnotesize mid} goes high, Q2 activates and match-line \deltext{ML}\addtext{ml} discharges. This is, therefore, an `OR match-line'. \deltext{This}\addtext{It} also explains why cue = $\overline{\text{cue}}$ = G{\footnotesize ND} yields `don't care'. The match-line only stays high when every \deltext{pixel}\addtext{CAM cell} in its \deltext{column}\addtext{data chain} registers no `\addtext{data-}miss'. 

\begin{figure}[!tb]
  \centering
  \includegraphics[width=0.48\textwidth]{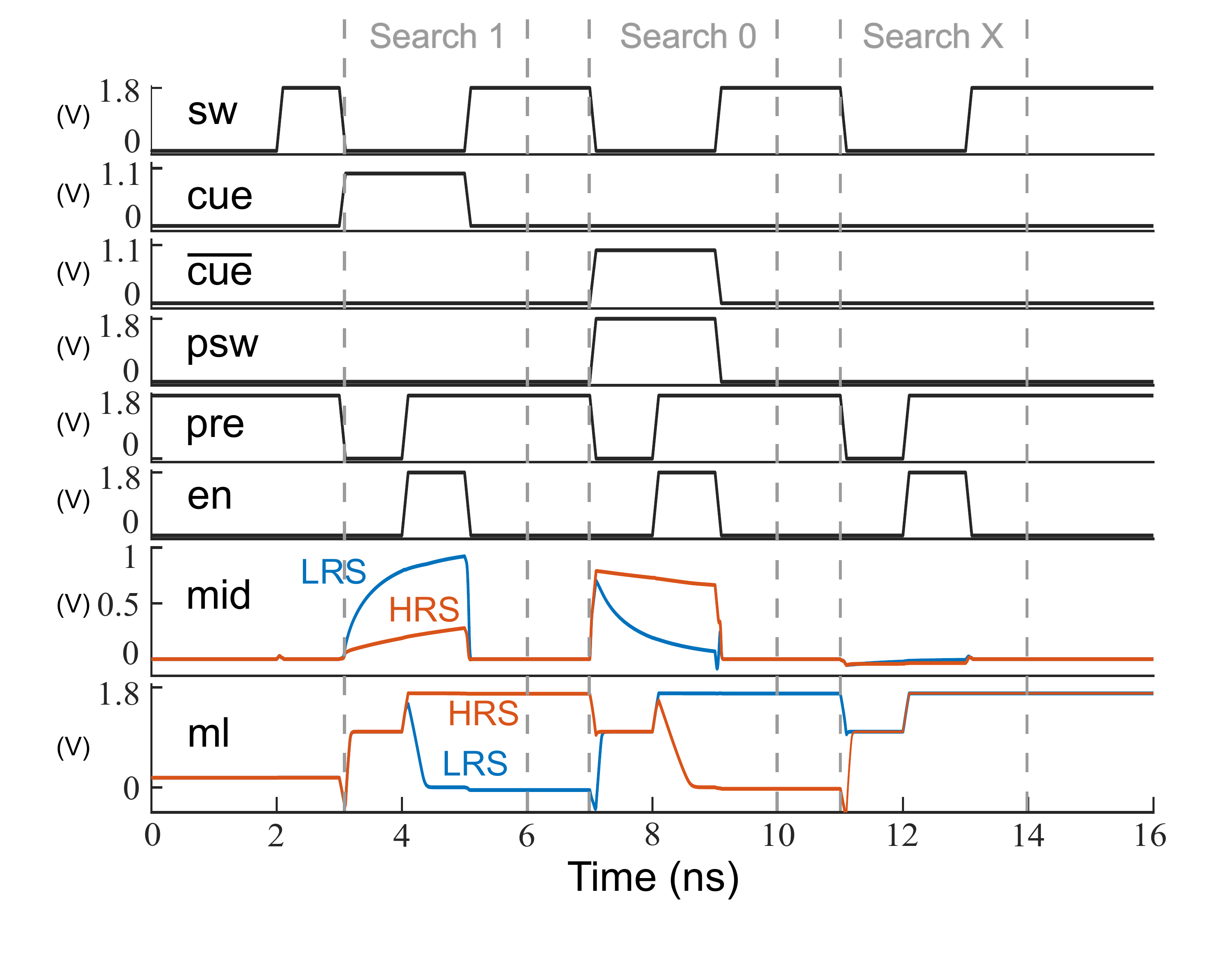}  
  \caption{CAR operation waveform for controlling searches and the corresponding outputs at match-line when RRAM are at LRS and HRS.}
  \label{fig:car_wave}
\end{figure}


Fig. \ref{fig:car_wave} demonstrates the CAR sequence in detail for \deltext{3x}\addtext{three} successive search operations. Before \addtext{and at the end of }each search, sw is \deltext{opened}\addtext{set to active high} to clear any residual charges at V{\footnotesize mid}. \deltext{Then}\addtext{At the first clock cycle of the CAR operation,} the match-line ml is pre-charged by the signal pre (active low) to the supply voltage level V{\footnotesize SEC} \deltext{for 1x clock cycle}. At the same time, cue and $\overline{\text{cue}}$ start to rise for 2ns according to the cue data. \addtext{During the rising of cue and $\overline{\text{cue}}$,} V{\footnotesize mid} starts to accumulate charges and its equivalent voltage exceeds the threshold voltage of Q2 only when a mismatch between the input data and the stored data is detected. \addtext{The psw is also set the same as $\overline{\text{cue}}$ to boost the bottom capacitor ratio. }\deltext{Then}\addtext{At the second clock cycle}, the signal en at the gate of Q3 is strobed when the value of V{\footnotesize mid} has stabilized sufficiently. Finally, voltage levels at the match-line (ml) at the \deltext{end of the phase}\addtext{third clock cycle} indicate the comparison result\addtext{s} \addtext{at 6ns, 10ns, and 14ns respectively}. \addtext{In this case, the delays of the CAM cell are always fixed and only depend on the clock frequency and number of clock cycles to complete an operation. Thus, the match-line result always remains and is captured at the end of operations.} Multi-bit data can be checked by chaining multiple \deltext{pixels}\addtext{CAM cells} with the same match-line. Every bit miss degrades the ml voltage level so that a percentage of hit/miss can also be observed given sufficiently sensitive sensing circuitry although here we restrict ourselves to binary answer detection.

\subsection*{Address-Addressable Read}
\begin{figure}[!htb]
  \centering
  \includegraphics[width=0.28\textwidth]{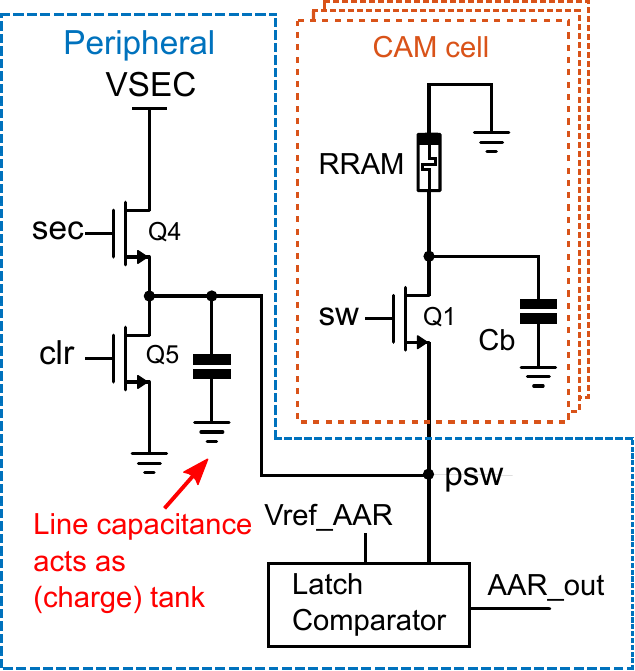}  
  \caption{AAR equivalent schematic for activating circuit part that produces AAR results. \deltext{Pixels}\addtext{The CAM cells} share the connection at psw node by accumulating charges at the line capacitance, and the output is augmented by a latch comparator.}
  \label{fig:aar_sch}
\end{figure}

\begin{figure}[!tb]
  \centering
  \includegraphics[width=0.32\textwidth]{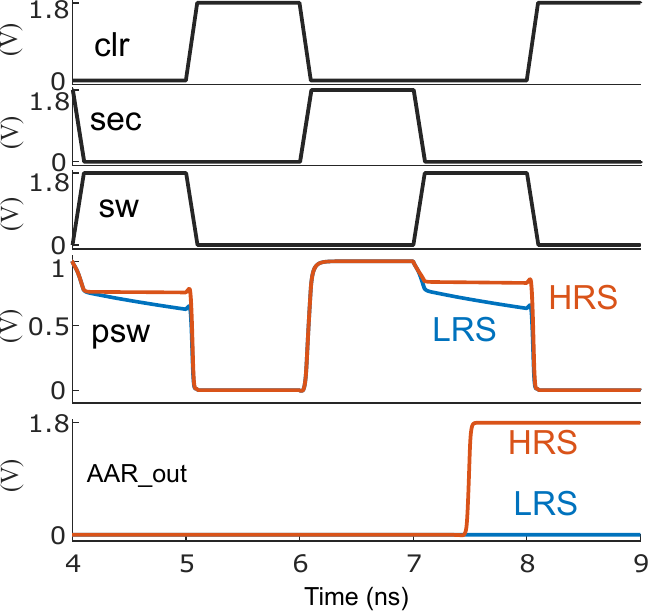}  
  \caption{AAR operation waveform with peripheral controlling input sequence and output obtained at the latch comparator when RRAM is at LRS and HRS.}
  \label{fig:aar_wave}
\end{figure}

\addtext{In data management systems, data manipulation does not only require the ability to search content in the memory but also retains access to the memory content by input addresses like the traditional SRAM/DRAM systems. The latter method is also called address-addressable read.} To enable access to the data bit that is stored in the \deltext{pixel}\addtext{CAM cell}, \addtext{only} the 1T1R structure (Q1 and RRAM) is activated, and the other part of the \deltext{pixel}\addtext{CAM cell} remains off. The equivalent circuit for the AAR operation is highlighted in \addtext{F}\deltext{f}ig. \ref{fig:aar_sch}. Both cue and $\overline{\text{cue}}$ are at G{\footnotesize ND}. \deltext{Net psw is connected to a horizontal chain of pixels (in array configuration).}\addtext{Numbers of CAM cells share the connection to the net psw in array configuration. Therefore, under} \deltext{In} this construction, the parasitic capacitance of the \addtext{psw} line \deltext{on psw creates}\addtext{can be regarded as} a charge tank to reflect the equivalent charge level when RRAM is at different states. The voltage levels at psw are reflected by latch comparators with a reference voltage Vref\_AAR. The operation waveform is shown in \addtext{F}\deltext{f}ig. \ref{fig:aar_wave}. The psw is firstly cleared by signal clr to open Q5 \addtext{at 5ns}. Next, a 1ns \deltext{strobing}\addtext{pulse} at the net sec sets psw to V{\footnotesize SEC}. Finally, sw is \deltext{strobed}\addtext{activated at 7ns} and \addtext{it} allows the line capacitance to discharge depending on the resistive state of the RRAM being read. Then the AAR results get amplified at AAR\_out. Only \addtext{one}\deltext{1x} \deltext{column of the array}\addtext{CAM cell on the same psw line} is activated in this configuration and sensing circuitry at the periphery determines the \addtext{AAR} result\addtext{s}.

\begin{figure}[!htb]
  \centering
  \includegraphics[width=0.18\textwidth]{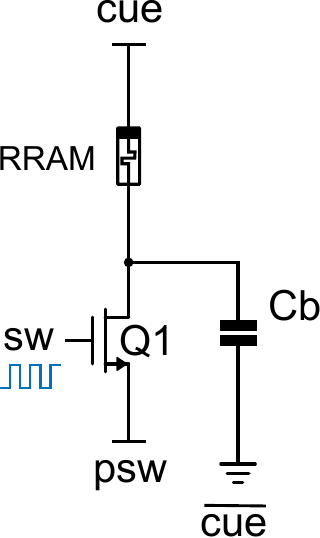}  
  \caption{Equivalent circuit schematic for the RRAM write operation and for the assisted ones that help to maintain higher voltages over RRAM.}
  \label{fig:wrt_sch}
\end{figure}

\begin{figure}[!tb]
  \centering
  \includegraphics[width=0.4\textwidth]{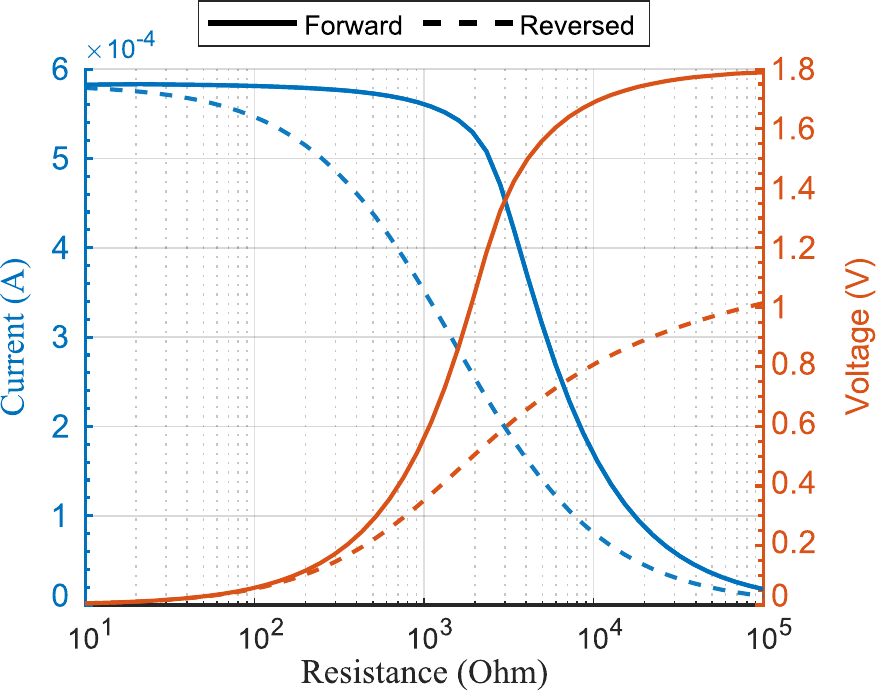}  
  \caption{Voltages and currents measured across the resistor sweeping from [10 to 100k]\textOmega \space during write operations in the forward direction and reversed direction.} 
  \label{fig:writeIV}
\end{figure}

\begin{figure*}[bt]
    \begin{minipage}[t]{.66\textwidth}
        \centering
        \includegraphics[width=\textwidth]{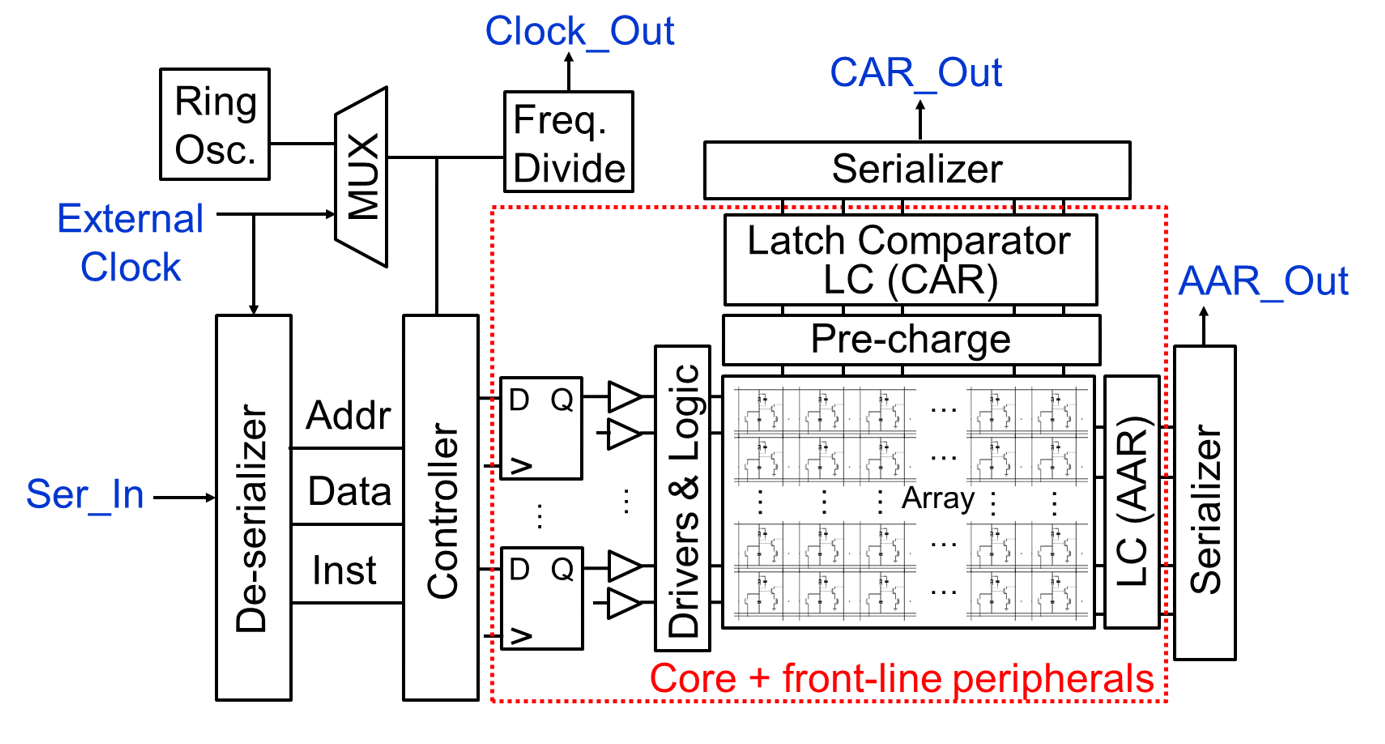}
        \subcaption{}\label{fig:sys}
    \end{minipage}
    \begin{minipage}[t]{.33\textwidth}
        \centering
        \includegraphics[width=\textwidth]{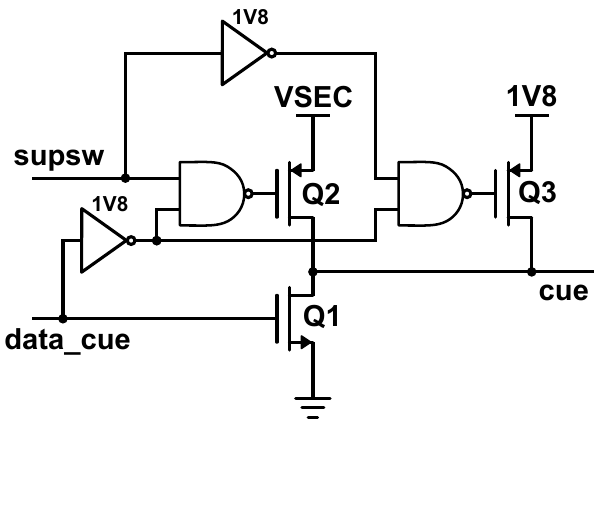}
        \subcaption{}\label{fig:cue_peri}
    \end{minipage}
    \begin{minipage}[t]{.4\textwidth}
        \centering
        \hspace*{2ex}\includegraphics[width=\textwidth]{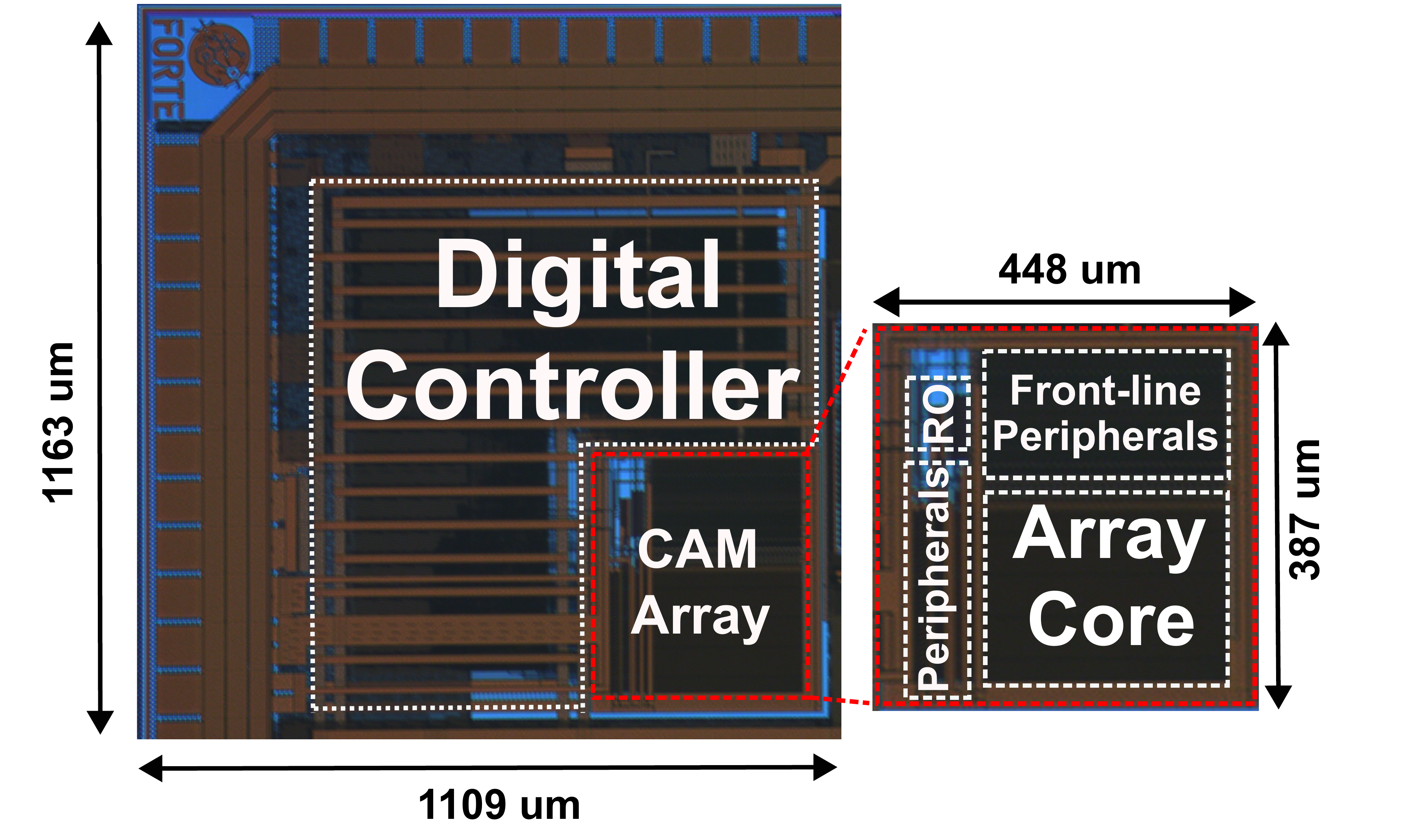} 
        \subcaption{}\label{fig:layout2}
    \end{minipage}
    \begin{minipage}[t]{.6\textwidth}
        \centering
        \hspace*{3ex}\includegraphics[width=\textwidth]{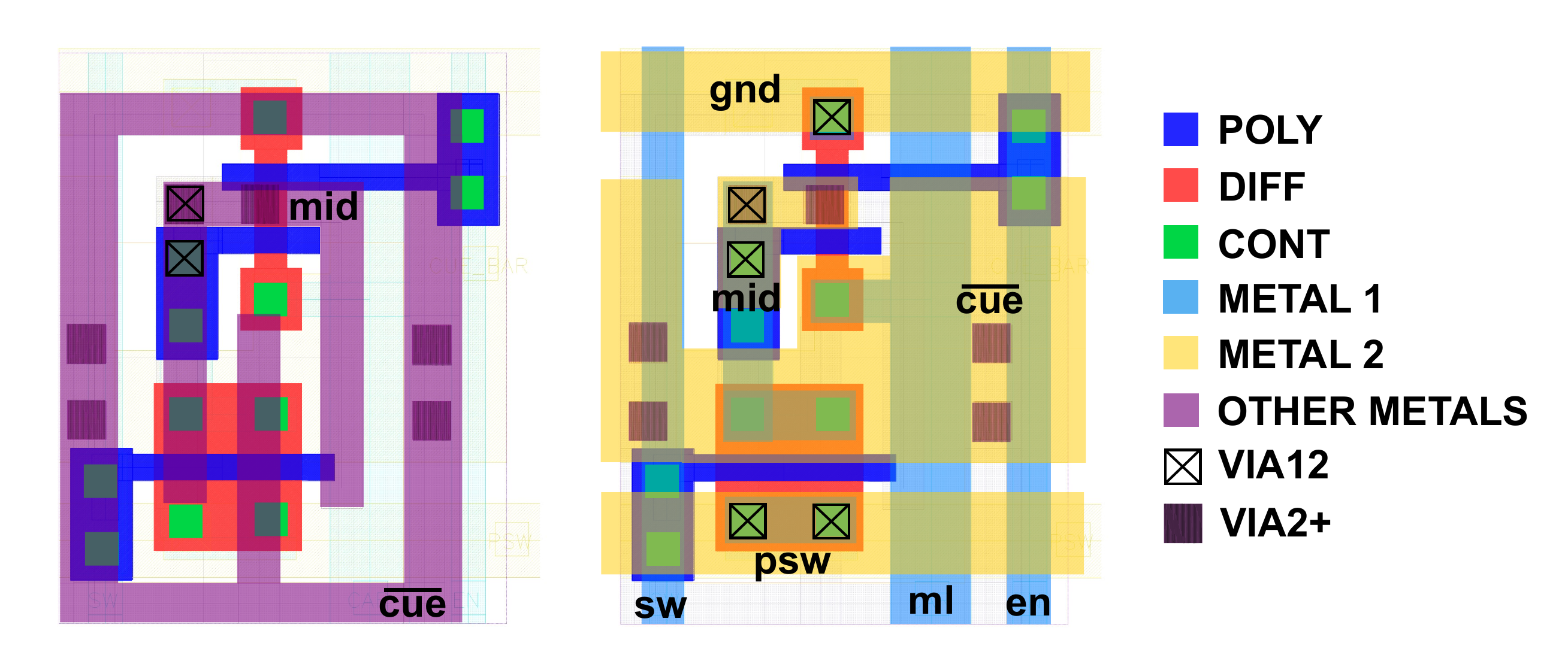} 
        \subcaption{}\label{fig:layout}
    \end{minipage}
    \caption{(a) System block diagram with IO whose core and front-line peripherals are fully customized. (b) Peripheral circuit for processing input data cue switching between read modes and writing modes. (c) The fabricated chip from the foundry without RRAM being post-processed. (d) \deltext{Pixel}\addtext{CAM cell} layout views. To improve visibility, the left shows the customized MOM capacitor placed above the transistors, and the right shows the remaining metal routes in the same \deltext{pixel}\addtext{CAM cell} without the capacitor.}
\end{figure*}

\subsection*{Write Operation}
The emerging RRAM device models used in this work assume electroforming-free operation. Programming is carried out using pulses. Fig. \ref{fig:wrt_sch} shows the activated \deltext{pixel }components \addtext{in the 3T1R1C cell} for the write operation. To write the RRAM to its desired state, cue and psw are the two terminals where the bias voltages are applied. We define forward direct writing when a bias voltage is applied at cue, and psw is connected to G{\footnotesize ND}. A reverse direction write is achieved by setting psw to V{\footnotesize DD} and keeping cue grounded. After biasing, sw is pulsed resulting in voltage applied across the RRAM. For a standard write operation, $\overline{\text{cue}}$ is always kept at G{\footnotesize ND} for both write directions. However, with the placement of C{\footnotesize b}, some system instantiations can be built to boost the voltages between the devices given the right conditions, but it is out of the scope of the present work. 

In principle, the RRAM writing conditions would be found by solving divider equations in each case. However, for simplicity, we sweep a linear resistor to observe the equivalent static resistance (ESR) at the DC operating point during the write operations. Fig. \ref{fig:writeIV} provides the IV across the ESR \deltext{in the pixel }in both forward and reverse directions during the write operations. The power supply is set to 1.8V. The forward direction allows more electric portions into the resistor as we have direct control of $V_{gs}$ but we get the source-degeneration effect in the reversed direction.

\subsection{System} \label{sec:system}
Fig. \ref{fig:sys} provides an overview of a 64x64 array of CAM \deltext{pixels}\addtext{cells}. \addtext{Each 64-bit of d}\deltext{D}ata is stored in columns since match lines are shared column-wise. Therefore, \addtext{the control signals for CAR operations} pre, en, and sw \addtext{are} also share\addtext{d} column-wise \deltext{so that the connected match-line can provide an equivalent voltage level indicating comparison similarity}\addtext{to make each column independently controlled}. The cue and $\overline{\text{cue}}$ are connected horizontally \addtext{for injecting the same set of search cue to all columns}. Signal psw is also connected horizontally to enable AAR operations for each row. In this way, the array can detect 64 sets of data with a 64-bit length. To analyze the outcome of CAR and AAR operations, two pMOS-type latch comparators are designed to resolve the potentially very small gap between the `all-hit' and `1-bit miss' scenarios for CAR and the resulting differences between HRS- and LRS-induced outcomes at psw for AAR. The cue is operated at the secondary adjustable power supply V{\footnotesize SEC} for CAR operation but it requires reaching the primary power supply V{\footnotesize DD} for WRT operations. Additionally, the signal at psw should be able to switch between the multiple supply voltages and the ground under different operation conditions. This is achieved via \addtext{three}\deltext{3x} transistors, with pri connecting to the primary 1.8V supply and sec to V{\footnotesize SEC}. As for $\overline{\text{cue}}$, it operates at V{\footnotesize SEC} for CAR but stays at the V{\footnotesize DD} or G{\footnotesize ND} for the other operations. The peripheral circuit for supplying cue is shown in \addtext{F}\deltext{f}ig. \ref{fig:cue_peri} where an additional signal supsw is introduced to swap between the supplies. To enable parallel CAR searching, each row is provided with its own peripheral circuits and signal drivers so that the entire array can be activated at the same time. These parts of the system are circled as front-line peripherals where they are implemented in the majority of CAM systems.

At the back-line peripherals that are not common to all systems, the chip receives inputs through a de-serializer, which splits the incoming information into address, data, and instructions. The finite state machine recognizes this and generates all control signals for the array. Those signals are gated by a group of flip-flops for synchronization and then delivered to the drivers and peripherals of the array. The outputs from CAR and AAR operations are sent to their corresponding serializers and output to the chip. For this work, we implemented an internal clock running at 875MHz to orchestrate the control signals but the system is also capable of operating at lower clock speeds by connecting to an external clock. 

\subsection{Layout}
The CAM array is implemented in 0.18um technology. Fig. \ref{fig:layout2} shows the fabricated chip that has been received from the foundry without in-house post-processing RRAM placed. The chip has an area of 1163um x 1109um where the digital ASIC area occupancy is not optimized. The \deltext{pixel}\addtext{CAM cell} layout is given in \addtext{F}\deltext{f}ig. \ref{fig:layout} with a dimension of 3.04um x 3.875um. The transistor Q2 and Q3 share one of the terminals as a double-gated transistor. To visualize \deltext{pixel}\addtext{the CAM cell} layout in layers, only the bottom customized MOM capacitor C{\footnotesize b} and \addtext{the three} transistors \deltext{in the pixel} are shown on the left of \addtext{F}\deltext{f}ig. \ref{fig:layout}. The other routing layers are presented on the right. The \deltext{pixel}\addtext{CAM cell} area is limited by the area of the bottom customized MOM capacitor C{\footnotesize b} which uses most of the metal layers, excluding METAL 1 and METAL 2 layers. \addtext{We note that this customized capacitor has an inter-digitated shape where the terminal connected to the net mid is encased into the terminal connected to $\overline{\text{cue}}$. This is no accident: the mid terminal, as the most noise-sensitive node, is surrounded and thus isolated by the $\overline{\text{cue}}$ terminal to avoid unwanted capacitance coupling to adjacent cells or indeed any other node. In contrast, node $\overline{\text{cue}}$ is actively driven, and thus charge-sharing with parasitic capacitors does not materially affect the voltages it assumes during operation.} The area of C{\footnotesize b} can be significantly decreased if more metal layers are available at a more advanced technology node. The same-size capacitance can be implemented by a smaller area so that the \deltext{pixel}\addtext{CAM bitcell} limiting factor becomes the size of the three transistors. 

\section{Measurements}
Due to the \deltext{development of} RRAM post-processing, this design has a limited number of metal layers available for routing. It also has a major impact on the customized MOM capacitor C{\footnotesize b} whose capacitance value can be greater if more metal layers are used within the same area. Thus, the simulations in the work are established with an additional bottom RRAM which is always in its pristine state with large resistance in Mega or Giga Ohms. There is no fundamental difference to the concept but it acts as an adjustment to the engineering realities of our particular test run. This is implemented for a higher C{\footnotesize b} to enlarge the capacitive divider ratio for testing strategy. It brings flexibility to altering C{\footnotesize b} when testing the chip as this second RRAM can be post-processed with different dimensions (different sizes of capacitance).

\subsection{Function and Timing}
According to the system implementation, the functionality test for the CAM array should be focused on the content searching of 64-bit data where 64 \deltext{pixels}\addtext{CAM cells} share the same match-line. The match-line should be able to distinguish between the worst-case 1-bit data miss and all data hit. The voltage differences read at the match-line are used to set the correct reference voltages for the latch comparator. 64-bit data has $2^{64}$ sets of combinations. An accurate function should be able to discriminate all combinations without adjusting the voltage reference of the latch comparator. Therefore, four types of combinations of RRAM resistance states (data) are considered: (1) 64HRS (2) 64LRS (3) 63HRS+1LRS (4) 1HRS+63LRS. At the same time, the searching cue is also set to these four cases. Their corresponding relations are listed in table \ref{tab:HitMiss} where all data hit (blue) and worst-case 1-bit data miss (red) are highlighted. Each case reflects the search result by output voltage levels at its match line. To identify hits from any misses, the output voltage levels for all data hits must have higher values than that for any worst-case 1-bit data miss. We define a threshold voltage between the lowest voltage for all data hits and the highest voltage for worst-case 1-bit data misses among the test cases in table \ref{tab:HitMiss}.

\begin{table}[!b]
\caption{Functional test cases for finding gaps between all data hit (blue) and worst-case 1-bit data miss (red).}
\label{tab:HitMiss}
\resizebox{\columnwidth}{!}{%
\begin{tabular}{@{}ccccc@{}}
\toprule
\multirow{2}{*}{\textbf{Search (cue)}} & \multicolumn{4}{c}{\textbf{RRAM Resistance State (data)}}   \\ \cmidrule(l){2-5} 
                        & \textbf{64HRS}        & \textbf{64LRS}       & \textbf{63HRS+1LRS} & \textbf{1HRS+63LRS} \\ \midrule
\textbf{\addtext{(1)} 64HRS}         & \textcolor{blue}{Hit}  & Miss                  & \textcolor{red}{Miss}   & Miss                    \\ \midrule
\textbf{\addtext{(2)} 64LRS}         & Miss                   & \textcolor{blue}{Hit} & Miss                    & \textcolor{red}{Miss}   \\ \midrule
\textbf{\addtext{(3)} 63HRS+1LRS} & \textcolor{red}{Miss}  & Miss                  & \textcolor{blue}{Hit}   & Miss                    \\ \midrule
\textbf{\addtext{(4)} 1HRS+63LRS} & Miss                   & \textcolor{red}{Miss} & Miss                    & \textcolor{blue}{Hit}   \\ \bottomrule
\end{tabular}%
}
\end{table}

\begin{figure}[!b]
  \centering
  \includegraphics[width=0.48\textwidth]{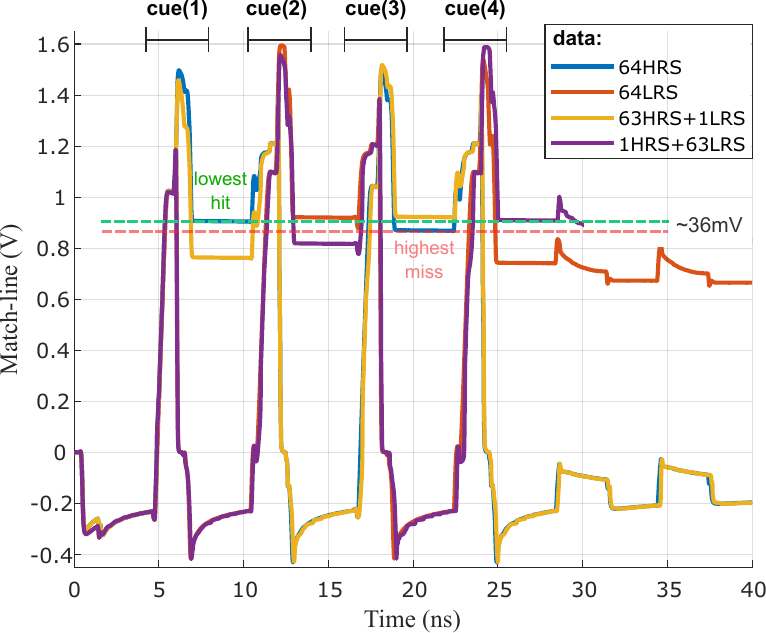}  
  \caption{Match-line equivalent voltages with four groups of RRAM (data) response to four types of searching (cue) (1) 64HRS, (2) 64LRS, (3) 63HRS+1LRS, and (4) 1HRS+63LRS. It reports a 36mV gap between the lowest data hit and the highest data miss. } 
  \label{fig:func4case}
\end{figure}

\begin{figure*}[!tb]
    \begin{minipage}[t]{.46\textwidth}
        \centering
        \hspace*{5ex}\includegraphics[width=\textwidth]{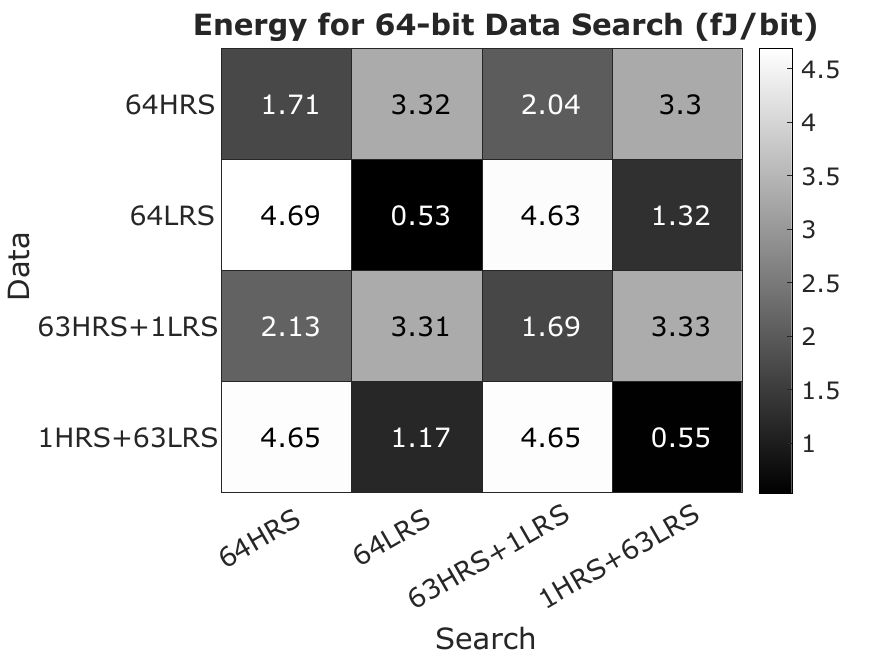}
        \subcaption{}\label{fig:power_map}
    \end{minipage}
    \hfill
    \begin{minipage}[t]{.46\textwidth}
        \centering
        \includegraphics[width=\textwidth]{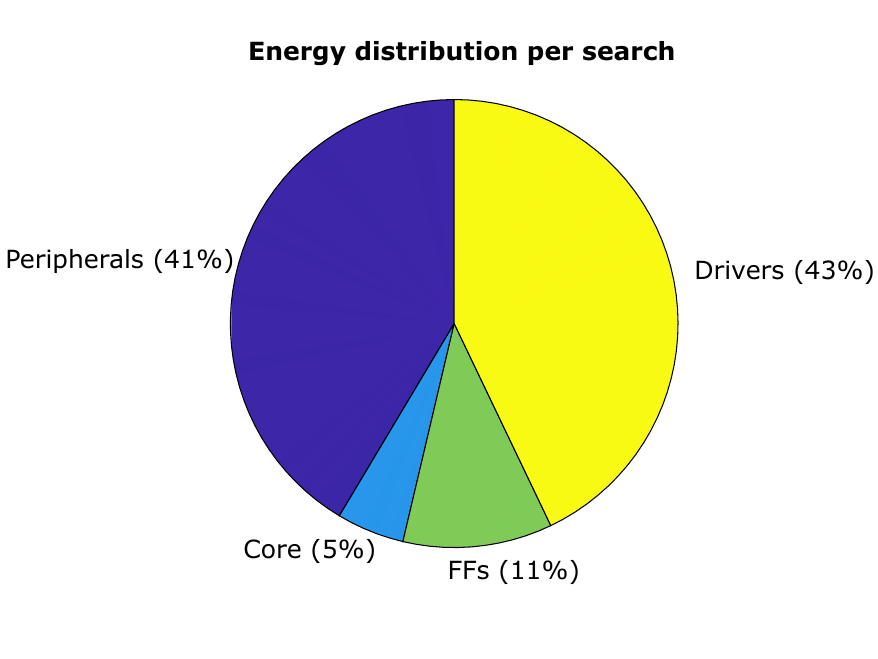}  
        \subcaption{}\label{fig:power_bar}
    \end{minipage}
    \caption{(a) Energy consumption per bit for 64-bit data under four test cases to define the best and the worst searching scenario. The energy measurement takes account of the core and logic peripherals in \addtext{F}\deltext{f}ig. \ref{fig:sys}. (b) Energy consumption breakdown for core and front-line peripherals.}
\end{figure*}

\addtext{A single CAM cell has two searching scenarios, looking for 1 or 0, thus there are HRS and LRS hits. They result in unequal voltages observed at the match line although both are data hits. The search cue (1) phase covers the 64 HRS hits and the worst-case on one HRS miss when searching for 1, while the search cue (2) phase gives the voltage levels at 64 LRS hits and the worst-case on an LRS miss when searching for 0. On the other hand, the worst case for an LRS miss when it is mostly searching for 1 is recorded in the search cue (3), and the search cue (4) tests for the worst case that an HRS miss when it is mostly searching for 0. With the pre-defined four cases, the system covers the full range of possible outcomes that 64-bit data may produce.}

Fig. \ref{fig:func4case} displays output voltages observed at match-line with the defined test sets for 64-bit data when V{\footnotesize SEC} is at 1.18V. The graph shows four groups of searching: 64HRS (5-8ns), 64LRS (11-14ns), 63HRS+1LRS (17-20ns), and 1HRS+63LRS (23-26ns), whose match-line is shared between 64 \deltext{pixels}\addtext{CAM cells}. The match line voltages with the four groups of RRAM resistance state arrangements reflect the results of the searches. For example, considering the blue trace, where there are 64 HRS RRAM, the first search gives a data hit and generates the highest equivalent voltage at 905.25mV in the cue(1) search phase. This RRAM trace also reaches an equivalent voltage of 869.07mV at the third search, representing a worst-case 1-bit data miss when searching for 63HRS+1LRS.\deltext{A single pixel has two scenarios of searching, looking for 1 or 0. Therefore, it has HRS hits and LRS hits that result in unequal voltages at the match line. The search cue (1) phase covers the 64 HRS hits and the worst-case on one HRS miss when searching for 1, while the search cue (2) phase gives the voltage levels at 64 LRS hits and the worst-case on an LRS miss when searching for 0. On the other hand, the worst case for an LRS miss when it is mostly searching for 1 is recorded in the search cue (3), and the search cue (4) tests for the worst case that an HRS miss when it is mostly searching for 0. With the pre-defined four cases, the system covers the full range of possible outcomes that 64-bit data may produce.} By obtaining the lowest data hit and highest worst-case 1-bit data miss, this test indicates a gap of 36.18mV between data hit and miss. The reference voltage of the latch comparator is then set at a threshold within the specified range to identify the results.

A positive equivalent voltage gap at the match-line represents a valid operating window to cover the full range of 64-bit data. A higher gap can release the constraints on readout peripherals. At the same time, this valid gap can be calibrated by adjusting the secondary supply V{\footnotesize SEC} which controls the voltage at mid directly. This mid voltage controls the gate of the transistor to pull down the match-line. With this flexible secondary supply voltage control, the circuit can adapt to speed, corners, and other operating conditions, detailed in the later sections.

\addRev{In terms of timing, we define match-line developing delay as the evaluation time for pre-charged match-line to be fully discharged at data all-miss cases. The detected time for the 64x64 CAM array is found at 143ps when searching for HRS and 163ps for LRS. On the other hand, the search delay is determined at the time required from input to the time when output can be measured. Thus, the time is made up by the pre-charge time and match-line evaluation time that is 1.163ns in total.}

\subsection{Energy}

\begin{table}[!b]
\centering
\caption{\deltext{Pixel}\addtext{CAM cell} energy consumption during CAR operation measured in fJ. RRAM has LRS equal to 112k$\Omega$ \space and HRS is 8.04M$\Omega$.}
\label{tab:3t1r1cenergy}
\begin{tabular}{@{}ccccc@{}}
\toprule
\textbf{Search data} & \textbf{RRAM} & \textbf{Pre-charge} & \textbf{Enable} & \textbf{Total} \\ \midrule
\multirow{2}{*}{HRS}   & LRS           & 3.05         & 5.82        & 8.87           \\ \cmidrule(l){2-5} 
                     & HRS           & 3.04         & 2.41        & 5.45           \\ \midrule
\multirow{2}{*}{LRS}   & LRS           & 3.47         & 4.08        & 7.55           \\ \cmidrule(l){2-5} 
                     & HRS           & 2.37         & 3.41        & 5.78           \\ \bottomrule
\end{tabular}
\end{table}

Energy consumption for this work is highly dependent on the RRAM resistive states, the capacitance value of $C_{b}$, and the secondary supply voltage V{\footnotesize SEC}. The RRAM resistive state determines the \deltext{conductivity of the pixel}\addtext{equivalent voltage} that is linked to the match-line pull-down strength. As for the capacitance, its energy has a relation of $E \propto CV^2/2$ where the voltage is supplied from V{\footnotesize SEC}. The system benefits from the capacitance structure where there is no direct path \deltext{in the pixel} during operations and therefore both idling and operating energy are cut to the minimum levels.

Table \ref{tab:3t1r1cenergy} summarized the energy consumption at the \deltext{pixel}\addtext{bitcell} level undertaken with a 4fF $C_{b}$. The circuit operated with a 1.8V power supply but the input data are at 1V. The energy is breaking down into CAR operating phases \addtext{including}\deltext{, (PRE for} pre-charging match-line and \deltext{EN}\addtext{enabling bitcell to}\deltext{for} release charges \deltext{to allow}\addtext{for} output evaluation. It is observed that the energy consumption is higher when the RRAM resistive state is at LRS. The energy consumption per \deltext{pixel}\addtext{CAM cell} reaches the highest at the searching miss between HRS in searching data and LRS in RRAM in \deltext{pixel}\addtext{the CAM cell}. This is because the match-line discharge transistor (Q2 in \addtext{F}\deltext{f}ig. \ref{fig:Pixel}) is in the most conductive scenario.

The \deltext{pixels}\addtext{CAM cells} are tied to a 64x64 array and the energy consumption per match-line per bit search for the four test cases is recorded in \addtext{F}\deltext{f}ig. \ref{fig:power_map}. It is observed with a reduction of energy per bit when the \deltext{pixels}\addtext{CAM cells} are tied to an array structure and share the match line. The worst energy consumption is detected at 4.69fJ/bit when all the data (in 64 HRS) are different from the search cue (in 64 LRS). A 1-bit miss in \deltext{pixels}\addtext{CAM bitcell} creates a path to discharge the match-line where the energy is consumed. Therefore, the best energy consumption cases happen when the data (in 64 LRS) is the same as the search cue (in 64 LRS). At the same time, it is observed that searching in LRS consumes less energy than that in HRS. In other words, searching in LRS means supplying voltage from the bottom capacitor $C_{b}$ end where the RRAM accesses energy indirectly.

\addtext{Overall, the key trait of our proposed system in terms of energy consumption is that during read and search operations there are no DC paths, courtesy of the fully capacitive load that memory cells present to their drivers. This has two effects: (1) Energy saving as our operations approach the ideal $CV^2/2$ limit and (2) the little energy that is consumed is largely consumed by the drivers. The latter effect occurs because the capacitive divider merely ``borrows" charge from the line drivers cue/$\overline{\text{cue}}$ during operation (contrast this to the 3T1R design where the RRAM device dissipates significant energy by itself). Thus, power within the confines of the CAM cell is largely dissipated by Q2-Q3 in case of a ``miss", during which the accumulated charge on the match-line is dumped through them.} The system (core and front-line peripherals) energy consumption breakdown in Fig. \ref{fig:power_bar} shows that the array core only consumes 5\% of overall front-line energy consumption, whereas the heavy drivers and peripheral circuits consume the majority of energy. \deltext{This is because the searching mechanism in this system borrows energy from its peripherals to service the capacitors. It contrasts with regular DC approaches, changing the distribution of power dissipation between peripherals and the core.} 

\subsection{Corners}

\begin{figure}[!b]
  \centering
  \includegraphics[width=0.45\textwidth]{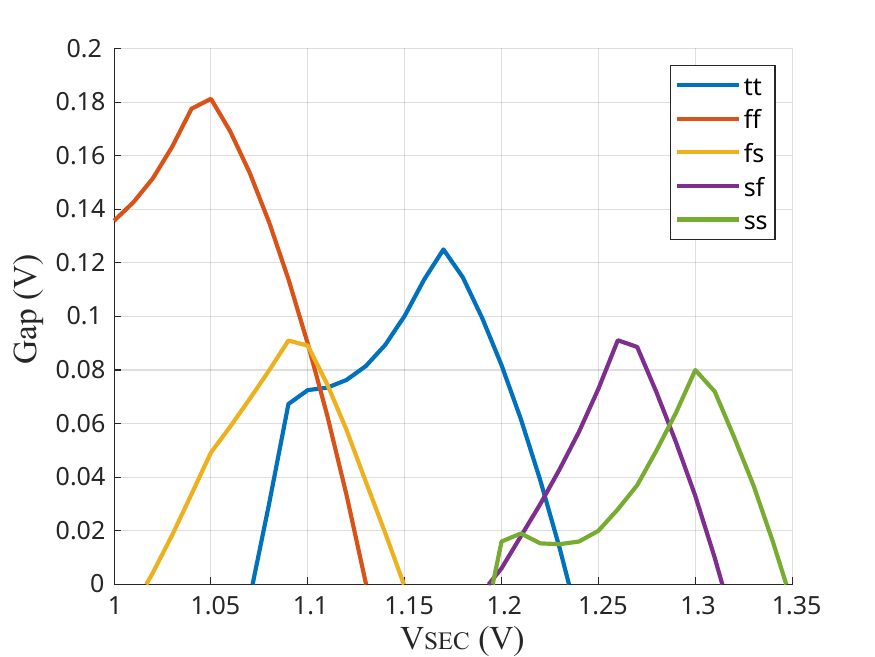} 
  \caption{Corner analysis for the 64x64 array core with pre-charge transistors, showing that the secondary power supply V{\footnotesize SEC} can be used to compensate for the PVT variations. The x-axis is sampled in steps of 10mV.}
  \label{fig:corner}
\end{figure}

\begin{table*}[!tb]
\centering
\caption{\addRev{TCAM performance comparison between CMOS-based TCAM and other non-volatile TCAMs}.}
\label{tab:compare}
\begin{threeparttable}
\resizebox{\textwidth}{!}{%
\begin{tabular}{lcccccccc}
\toprule
{  } &
  {12T CMOS \cite{12tcmos}} &
  {2T-2PCM \cite{li20132t2rIBM}} &
  {2FeFET \cite{2fefet}} &
  {2T-2MTJ \cite{2t2mtj}} &
  {3T-1RRAM \cite{chang20173t1r}} &
  {2T-2RRAM \cite{2t2r65nm}} &
  {This work} \\ \midrule
{Technology Node (nm)} &
  {28} &
  {90} &
  {45} &
  {28} &
  {90} &
  {65} &
  {180} \\ \midrule
{Non-volatility} &
  {No} &
  {Yes} &
  {Yes} &
  {Yes} &
  {Yes} &
  {Yes} &
  {Yes} \\ \midrule
{Word Size} &
  {64} &
  {64} &
  {64} &
  {32} &
  {64} &
 {64} &
  {64} \\ \midrule
{Frequency (MHz)} &
  {10} &
  {-} &
  {-} &
  {333} &
  {10} &
  {500} &
  {875} \\ \midrule
{Search Voltage (V)} &
  {1.00} &
  {1.20} &
 {0.60} &
  {1.20} &
  {1.00} &
  {0.70} &
  {1.80 \& 1.24} \\ \midrule
{Search Delay (ns)} &
  {1.34} &
  {1.90} &
  {0.40\tnote{1}} &
  {1.44\tnote{1}} &
  {0.96} &
  {0.58\tnote{1}} &
  {1.16} \\ \midrule
{Search Energy (fJ/bit/search)} &
  {1.62} &
  {-} &
  {0.116\tnote{2}} &
  {1.73} &
  {0.51} &
  {0.23\tnote{3}} &
  {4.15\tnote{4}} \\ \bottomrule
\end{tabular}%
}
  \begin{tablenotes}
    \item[1] \addRev{Match-line evaluation delay only.}
    \item[2] \addRev{Energy for pre-charge and sense amplifier included.}
    \item[3] \addRev{Energy for TCAM only.}
    \item[4] \addRev{Energy includes core and front-line peripherals for searching 64 entries of 64-bit in parallel.} 
  \end{tablenotes}
\end{threeparttable}
\end{table*}

To \deltext{observe the}\addtext{detect} search outcome\addtext{s}, a wider gap between the all-bit hit and worst case one-bit miss helps in releasing the stress of the readout circuit. In another aspect, \deltext{the}\addtext{a} wider gap means that more bits can be tied together \deltext{to the data}\addtext{resulting in longer data length}. T\addtext{he adjustable secondary supply V{\footnotesize SEC} can be adjusted to find t}his maximum gap operating point \addtext{under various conditions}.\deltext{can be adjusted by controlling the adjustable secondary supply V{\footnotesize SEC}.}

With the PVT variation, the system reflects the effect at the mid node equivalent voltages\addtext{,} leading to a substantial difference at the maximum gap operating point if the voltages are near the threshold voltage of transistors. Fig. \ref{fig:corner} plots the output gaps measured under the five corners from ff to ss with the variation of V{\footnotesize SEC} in 10mV sweeping from [1, 1.35V]. The results show that the maximum gap for ff corner is achieved when V{\footnotesize SEC} is at 1.05V and gives the highest gap among the tests. At the ss corner, a 1.3V V{\footnotesize SEC} is required to reach its maximum gap.


\section{Discussion}

%

In this work, the charge domain content addressable memory does not have a direct current path during searches, which \delRev{leads to a minimum energy consumption}\addRev{reduces DC power consumption. The circuit is benchmarked with a frequency of 875MHz and worst-case 1-bit data miss}. \delRev{At the same time, i}\addRev{I}t also functioned with address-addressable read which is also a charge-limited operation, albeit the charge limitation happens at the bit line level and involves the design of the peripherals. 
Table \ref{tab:compare} \addRev{presents the TCAM performance between CMOS-based memory and other non-volatile-based memories.}
\addRev{Overall, non-volatile TCAMs show relatively comparable performance in terms of cell area. The FeFET and RRAM TCAMs also improve search delay compared to 12T CMOS memory.}
\delRev{lists the energy consumption for various non-volatile TCAM designs, benchmarking the design on the basis of a 10x10 array at 0.18um technology node in [21] and the measurement for this work at the same technology node. Their energy per bit search is recorded in fJ/bit-search. The 2T2R and the 4T2R design uses two RRAMs and they require the RRAM to operate in three different resistive states. It also sets stringent constraints to the RRAM that can operate with these designs.}
\delRev{As for the 3T1R design, the bitcell requires a relatively low resistance for LRS, otherwise, the RC recovery time gets longer and therefore the operating speed is limited. Its direct current path also indicates a large amount of in-efficient energy consumption.} 
\addtext{This \addRev{table} provides an indicative comparison of performance, but it is important to note that \addRev{the search delay and energy measurements from literature are defined differently. In some works, search delay considers match-line evaluation time only, while other works also added pre-charge time. As word size increases, the pre-charge time can be a dominant portion of delay time. Therefore, delay time should be interpreted carefully during comparison. It also applies to search energy comparison which can be affected by data set used for benchmarking and the peripherals that are included. In this work, the search energy includes the necessary peripherals that are used to activate the 64x64 CAM array. It includes not only the pre-charge transistor and latch-comparators but also the heavy drivers for each control signal used to drive 64 bitcells. } 
\delRev{Other factors also affect the result: the required/used resistive state values for HRS and LRS, as well as the curvature of the IV characteristics will also impact performance and are difficult to control for.} 
\delRev{Table IV shows that our proposed approach is promising and worth closer investigation.}} 

\deltext{However, n}\addtext{N}on-volatile CAM arrays are \delRev{optimized with the non-volatile devices that they are used, whose}\addRev{tailored to the non-volatile devices they incorporate, with} IV characteristics \addRev{that} can directly influence the system's performance. 
\addRev{For RRAM-based CAMs, the required resistive state values for HRS and LRS and the curvature of the IV characteristics significantly affect performance and are challenging to unify for comparison.}
It is unfair to define the energy measurement with \addRev{only} \deltext{pixel}\addtext{bitcell} energy, \delRev{only }since the capacitive elements involve a flow of charge to its peripherals and may lead to more energy consumption measured at the peripherals. On the other hand, even testing with the same RRAM structure, the resistance value can be treated as a high resistance state in one design but would be regarded as a low resistance state in another circuit. The materials that are used to form the device determine the parasitic capacitance of the device and further affect the choice of the physical capacitor in this work. Both RRAM operating resistance states and capacitance in \deltext{pixel}\addtext{each bitcell} contribute to the major energy consumption of the array core. Therefore, it brings challenges to comparing the performances between CAM designs and the performance between different RRAM models in the same design. 

The designed CAM in this work can have 64 different data stored with a length of 64-bit. The searching ability of the array in terms of the data length is reflected by the length of the data chain (number of \deltext{pixels}\addtext{CAM cells} in a column). \deltext {This pixels}\addtext{These CAM cells} share a match-line and control the output gaps of the searched results. More \deltext{pixels}\addtext{CAM cells} make it harder to observe a difference between the lowest all-bit match and the highest one-bit miss cases. When the gaps become too small, an accurate sense amplifier that has high precision would required. At the same time, the searching capacity on the amount of data sets is affected by the \deltext{pixels}\addtext{number of CAM cells} in a row \deltext{which}\addtext{that} share the driver for each search cue. When more \deltext{pixels}\addtext{number of CAM cells} share the same driver, it means an increase in the driving strength as the overall capacitance per row increases. The system operates by a sequence of control\delRev{s}\addRev{ signals} that involve switching activities. These switching activities are directly linked to power consumption, especially for those large-sized drivers and peripherals. When the scale of data becomes large, performance optimization may target the way of sharing peripherals trade-off against speed.

This work is implemented in 0.18um technology whose digital ASIC part occupies at least three times greater than the RRAM CAM array. The system operating frequency is limited by the technology rather than the circuit. \addtext{On the other hand, the CMOS metal layers are mostly used for the customized MOM capacitor, leaving limited metal layers for routing.} With the scaling of technology, the chip density would be increased and the digital ASIC area could be shrunk. At the same time, the circuit can operated at a higher speed. The RRAM is assumed to have an area of 0.35um x 0.35um whereas, it can be fabricated with a smaller area of 0.1um x 0.1um, depending on the fabrication equipment and conditions. The area reflects the device's parasitic capacitance and further determines the choice of the physical capacitance \deltext{in pixels}as they form a capacitive divider. The secondary power supply V{\footnotesize SEC} helps with compensating such variants \deltext{in pixels}and makes the \deltext{pixels}\addtext{CAM cell} operate in the correct voltage range. \addtext{Overall, the major factors that can be improved with technology down-scaling are (1) energy consumption reduction due to the reduction of system power supply; (2) smaller design footprint due to more available metal layers (where available); (3) faster system operating frequency with the help of advanced standard cells from foundry; (4) smaller bottom capacitance with RRAM downscaling to a smaller area also benefit the system from energy and area perspective (enabling lower capacitances all around).}

Moreover, the system is insensitive to the operating frequency with the theory of performing RRAM RS as an open or short circuit structure. In other words, we define the LRS limit towards zero and the HRS limit towards infinite. In the HRS cases, the infinite resistance made the circuit follow the capacitive divider, while the LRS case made the circuit follow the input cue, where both cases are frequency-independent.

\addtext{Without CAM cells, the data searches in memory are normally achieved by reading data from each address and comparing them outside the memory. This process involves a large amount of data movements and is a bottleneck that constrains the system's performance. Moreover, with the growth of data sizes, d}\deltext{D}ata centers are becoming unsustainably power-hungry. \addtext{The}\deltext{such} \addtext{compact and} energy-efficient memory \addtext{proposed in this work benefits large databases in terms of speed and energy consumption}\deltext{is beneficial to these systems with large databases} with abilities to search data in parallel. \addtext{The combination of content-addressable read and address-addressable read means that the same hardware can be configured to operate either as a CAM or as a standard RAM, opening the way to greater operational flexibility. For example, it means that an enterprise storing its data in a standard relational database can transition to storing graph database information without changing hardware or even maintaining a hybrid database. The significance of this should not be underestimated.}\deltext{The implication of capacitance also brings possibilities of recycling the flow of charges in capacitors for further energy reduction. }

\section{Conclusion}

In conclusion, this paper presents a 3T1R1C capacitive-\deltext{memristive}\addtext{RRAM} TCAM array that uses RRAM as data storage and data comparison. This charge domain non-volatile CAM design applies a physical capacitor to form a capacitive divider with the RRAM parasitic capacitance, featuring no direct current paths and low leakages. The operating mechanism on 64-bit data and the method for finding its valid operating window are analyzed. The CAM supports both content-addressable read and address-addressable read with charge-limited operations. It reports an average energy of 64-bit data match at 1.71fJ/bit-search and 4.69fJ/bit-search for 64-bit data misses with an operating speed of 875MHz in 0.18um technology. With the continuing development of RRAM, we believe this CAM can be part of mainstream memories for powering future databases in data centers all around the world.

\section*{Acknowledgment}
This work has been supported by the Engineering and Physical Sciences Research Council (EPSRC) grants EP/V008242/1,2 and EP/R024642/1,2.

\bibliographystyle{IEEEtran}
\bibliography{IEEEabrv,TCAS_ASOCA1}

\begin{IEEEbiography}
[{\includegraphics[width=1in,height=1.25in,clip,keepaspectratio]{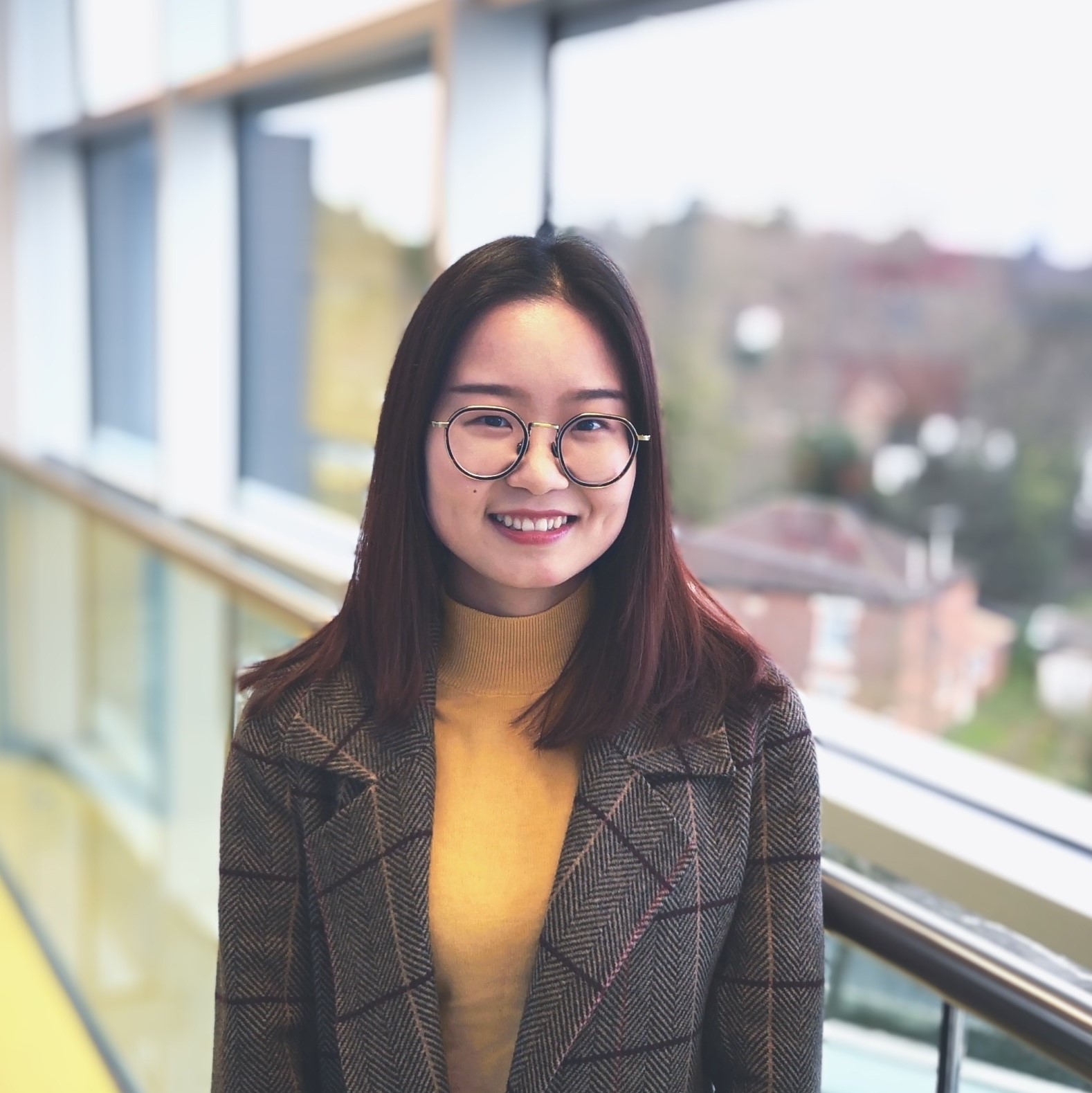}}]{Yihan Pan}
(Graduate Student Member, IEEE) received her B.Eng. degree in Electronic Engineering at the University of Manchester, U.K. in 2018 and her M.Sc. degree in Analogue and Digital Integrated Circuit Design at Imperial College London, U.K. in 2019. She is pursuing a Ph.D. degree at the Centre for Electronics Frontiers, Institute for Integrated Micro and Nano Systems, University of Edinburgh, U.K. Her research interests include hardware topologies for symbolic processing and RRAM-based memory architectures.
\end{IEEEbiography}

\begin{IEEEbiography}
[{\includegraphics[width=1in,height=1.25in,clip,keepaspectratio]{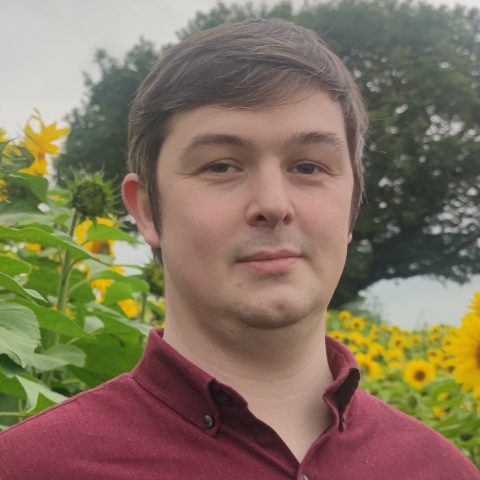}}]{Adrian Wheeldon}
(Member, IEEE) is a Research Associate at the Centre for Electronics Frontiers, University of Edinburgh. He received the M.Eng. degree in Electronic Engineering with Computer Systems from the University of Southampton in 2016, and the Ph.D degree in Electronic Engineering from the Newcastle University in 2022.
Adrian has developed demonstrator hardware for unconventional machine learning algorithms in both FPGA and ASIC. He has been involved in 3 ASIC tapeouts, holds several patents and has published 10+ papers in the areas of machine learning, vector symbolic architecture and digital circuit design.
\end{IEEEbiography}

\begin{IEEEbiography}
[{\includegraphics[width=1in,height=1.25in,clip,keepaspectratio]{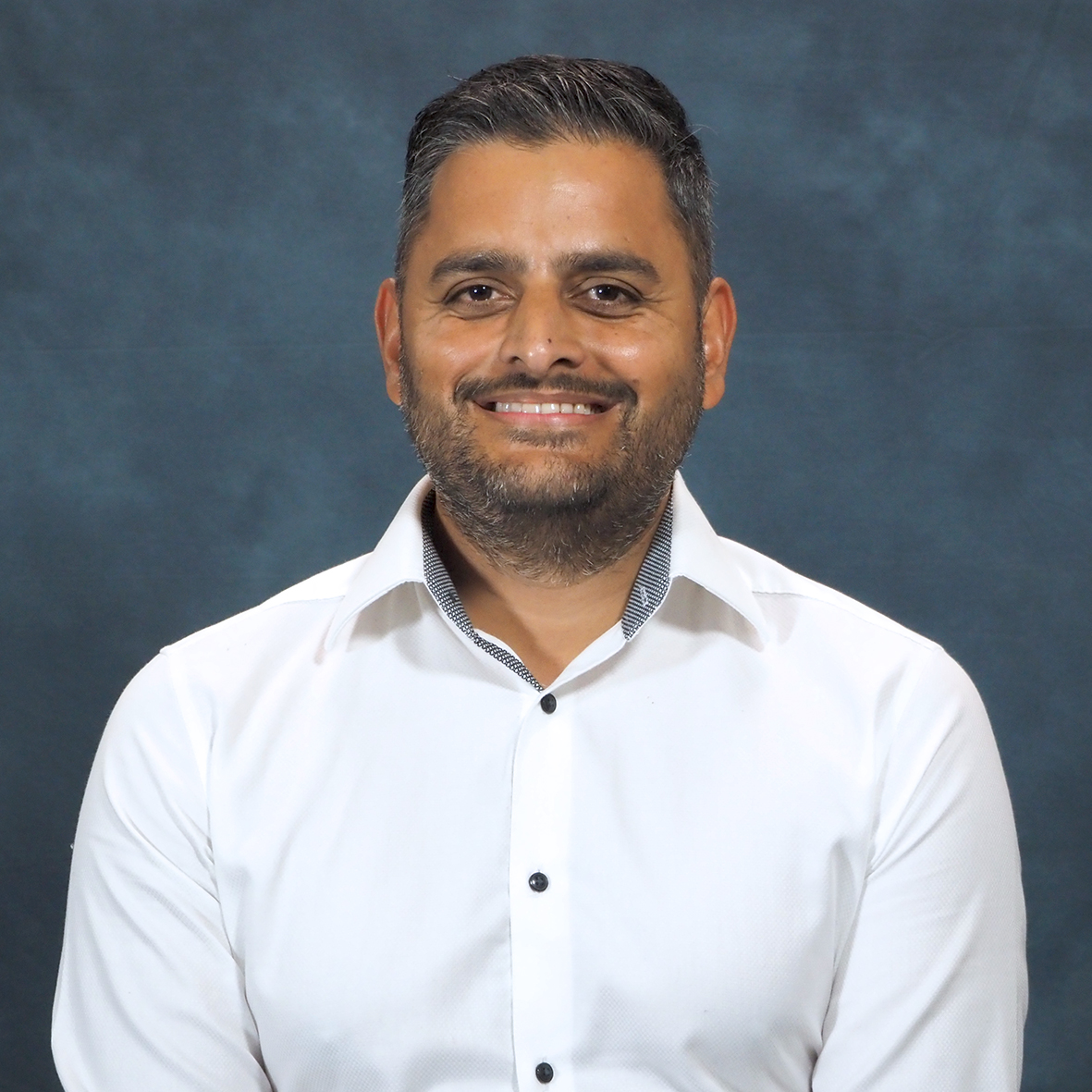}}]{Mohammed Mughal}
(Member, IEEE) is a Postdoctoral Research Associate at the meLAB, James Watt School of Engineering, University of Glasgow. He holds his Ph.D. and Postgraduate Diploma (PGD) from the University of Glasgow in 2017 and 2012 respectively. He worked as a KTP associate in flexible IC design, research fellow, and research associate before joining meLAB. His research interests include Analog and Mixed signal design, cryo-CMOS IC design for quantum computing, image sensors, analog neuromorphic computing, and AI hardware design.  
\end{IEEEbiography}

\begin{IEEEbiography}
[{\includegraphics[width=1in,height=1.25in,clip,keepaspectratio]{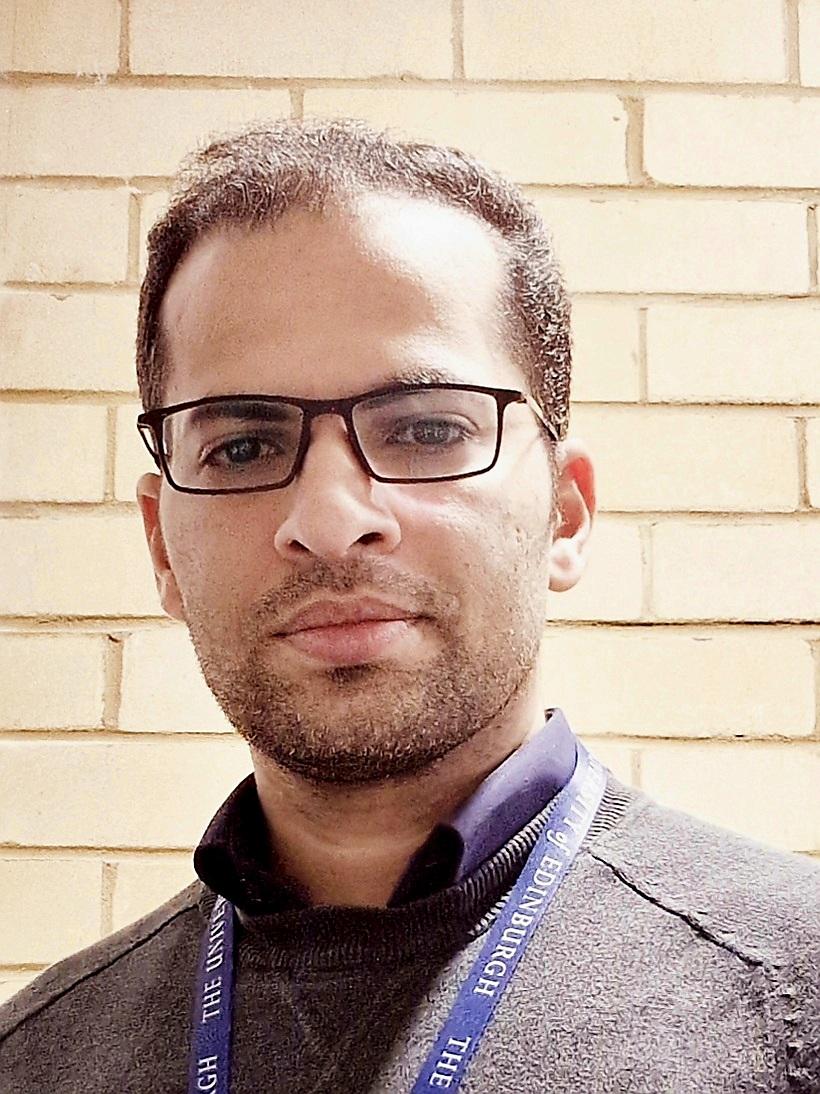}}]{Shady Agwa}
(Member, IEEE) is a Research Fellow at the Centre for Electronics Frontiers CEF, The University of Edinburgh (UK). He received his BSc and MSc degree from Assiut University (Egypt), both in Electrical Engineering. He got his PhD in Electronics Engineering from The American University in Cairo (Egypt) in 2018. Following his PhD, he joined the Computer Systems Laboratory at Cornell University (USA) as a Postdoctoral Associate for two years. In 2021, Shady joined the Centre for Electronics Frontiers at the University of Southampton (UK) as a Senior Research Fellow and then as a Research Fellow at the University of Edinburgh (UK). His research interests span across VLSI and Computer Architecture using conventional and emerging technologies for AI applications. His work focuses on unconventional ASIC-Driven AI Architectures which cover In-Memory Computing, Stochastic Computing, Systolic Arrays, Content-Addressable Memories, Beyond-Von Neumann Architectures and Energy-Efficient Digital ASIC Design.
\end{IEEEbiography}

\begin{IEEEbiography}[{\includegraphics[width=1in,height=1.25in,clip,keepaspectratio]{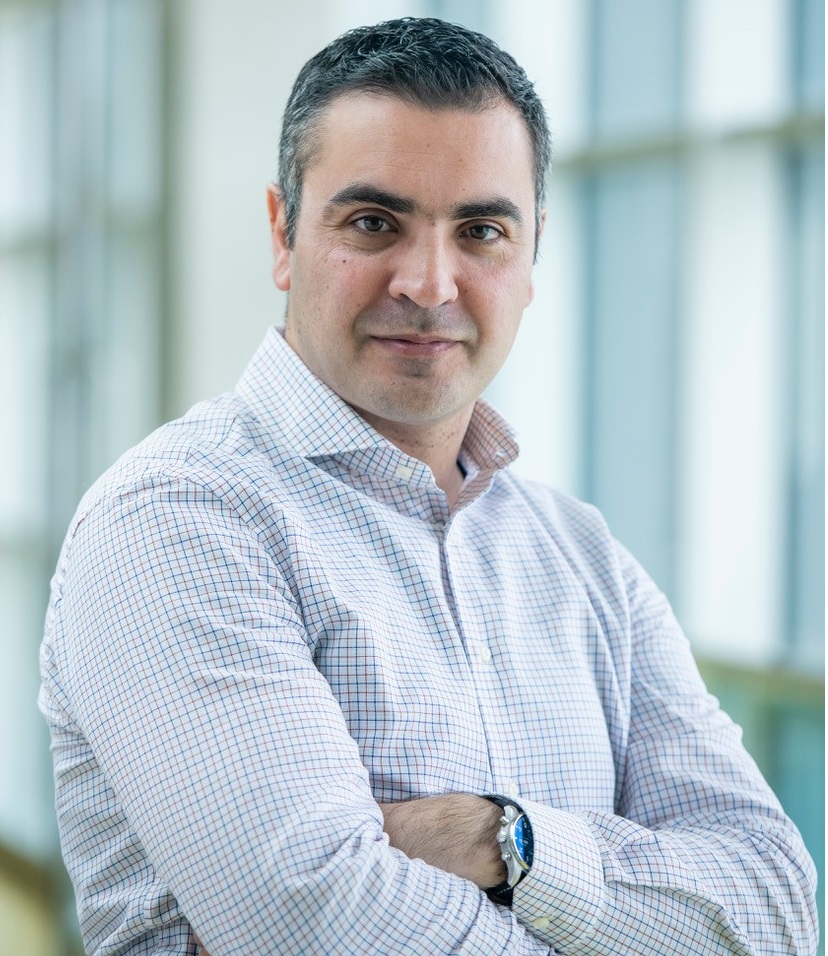}}]{Themis Prodromakis}
(Senior Member, IEEE) received the bachelor’s degree in electrical and electronic engineering from the University of Lincoln, U.K., the M.Sc. degree in microelectronics and telecommunications from the University of Liverpool, U.K., and the Ph.D. degree in electrical and electronic engineering from Imperial College London, U.K. He then held a Corrigan Fellowship in nanoscale technology and science with the Centre for Bio-Inspired Technology, Imperial College London, and a Lindemann Trust Visiting Fellowship with the Department of Electrical Engineering and Computer Sciences, University of California at Berkeley, USA. He was a Professor of nanotechnology at the University of Southampton, U.K. He holds the Regius Chair of Engineering at the University of Edinburgh and is Director of the Centre for Electronics Frontiers. He is currently a Royal Academy of Engineering Chair in emerging technologies and a Royal Society Industry Fellowship. His background is in electron devices and nanofabrication techniques. His current research interests include memristive technologies for advanced computing architectures and biomedical applications. He is a fellow of the Royal Society of Chemistry, the British Computer Society, the IET, and the Institute of Physics.
\end{IEEEbiography}

\begin{IEEEbiography}[{\includegraphics[width=1in,height=1.25in,clip,keepaspectratio]{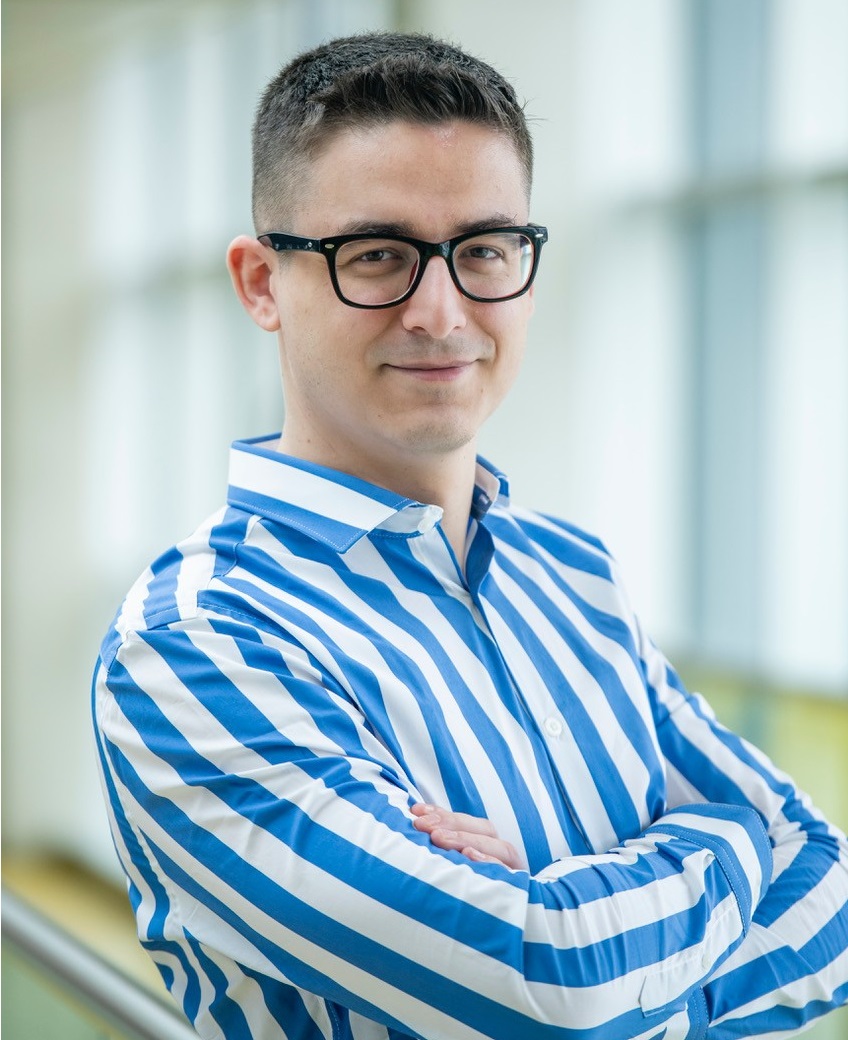}}]{Alexantrou Serb}
(Senior Member, IEEE) received the degree in biomedical engineering and the Ph.D. degree in electrical and electronics engineering from Imperial College in 2009 and 2013, respectively. He was a Research Fellow at the Zepler Institute (ZI), University of Southampton, U.K. He joined University of Edinburgh as a Reader in School of Engineering in 2022, where his research interests are cognitive computing, neuroinspired engineering, algorithms, and applications using RRAM, RRAM device modeling, and instrumentation design.
\end{IEEEbiography}
\vfill




\end{document}